\documentclass[12pt, centerh1]{article}
  \usepackage{metalogo} 
  \usepackage{algpseudocodex} 

  \usepackage{xcolor}
  \usepackage{array}
  \usepackage{soul}
  \usepackage{enumitem} 
  \usepackage{bm}
  \usepackage{amsmath}
  \usepackage{natbib}
  \usepackage{hyperref}
  \usepackage{amssymb}
  \usepackage{amsthm}
 \usepackage{subcaption}
\usepackage{float}
\usepackage{mathrsfs}
\usepackage{lineno}

 \usepackage{graphicx}
  \graphicspath{%
    {converted_graphics/}
    {Figures-2016-06-13/}
}

\newtheorem{theorem}{Theorem}[section]

\DeclareMathOperator{\argmax}{argmax}
\definecolor{jade}{rgb}{0.0, 0.66, 0.42} 
\newcommand{\com}[1]{\textcolor{black}{#1}}
\newcommand{\comR}[1]{\textcolor{black}{#1}}

  \title{Estimating Extreme Wave Surges in the Presence of Missing Data}

  \author{James H. McVittie\textsuperscript{1} and Orla A. Murphy\textsuperscript{2}*}
  \date{\small \textsuperscript{1}Department of Mathematics and Statistics, University of Regina, Saskatchewan, Canada\\
  \textsuperscript{2}Department of Mathematics and Statistics, Dalhousie University, Nova Scotia, Canada\\
  *Corresponding author. E-mail: orla.murphy@dal.ca}

\begin{document}
\maketitle
\begin{abstract}
The block maxima approach, which consists of dividing a series of observations into equal sized blocks to extract the block maxima, is commonly used for identifying and modelling extreme events using the generalized extreme value (GEV) distribution. In the analysis of coastal wave surge levels, the underlying data which generate the block maxima typically have missing observations. Consequently, the observed block maxima may not correspond to the true block maxima yielding biased estimates of the GEV distribution parameters. Various parametric modelling procedures are proposed to account for the presence of missing observations under a block maxima framework. The performance of these estimators is compared through an extensive simulation study and illustrated by an analysis of extreme wave surges in Atlantic Canada. \\ \vspace{0.5cm}

\end{abstract}
\newpage

\section{Introduction}
The analysis of extreme wave heights and risks of flooding using coastal wave surge data is typically fraught with missing observations. This missing data phenomenon is not restricted to coastal wave surge data but can occur in any extreme value analysis where the measuring apparati fails due to the intensity level of the measurements. Under the block maxima approach to extreme value analysis, which models the distribution of the maximum from a block of observations, missing observations at a series level (i.e. within blocks) is a common occurrence. However, it is generally unknown whether the missing observations are extreme or non-extreme. Consequently, whether the true block maxima have been observed or are missing cannot be determined. 

The analysis of extreme values in the presence of missing observations has been partially examined in the literature. Various authors including \cite{ayiah}, \cite{beirl}, \cite{xu} and \cite{zou} studied the Hill estimator when a subset of the higher order statistics were missing, whereas \cite{glava} studied the extremal properties of the moving average for incomplete data. In contrast, \cite{beck} proposed a general rule of thumb for ignoring missingness when fitting the generalized extreme value (GEV) distribution whereas \cite{turki2020} utilized imputation to handle the missing observations. Although the effects of missing higher order statistics within a block have been investigated, there is a paucity of work focused on accounting for missingness directly in the estimation of the unknown GEV distribution parameters  \citep{ryden,einmahl2008}. \comR{Other authors who have applied extreme value models in flood frequency analysis include \cite{jin}, \cite{kidson} and \cite{yan}.}

The primary purpose of this paper is to present a different methodological perspective to account for missing observations within the block maxima approach. Specifically, the assumptions on the missing data mechanism can be reinterpreted from the lens of the true block maximum being potentially right censored. That is, if a block of observations contains missing records, then it necessarily follows that the true block maximum is at least as large as the maximum recorded observation within the block (i.e. the observed block maximum). Through this view, it is apparent that missing observations within blocks can be connected to a right-censoring framework, from survival analysis, at the block level. Although the analysis of extreme values has been investigated for randomly censored data, these studies have primarily focused on estimation in cases of censoring at the series level rather than at the block level \citep{prescott1983, beirlant2010, stupfler2019}. Therefore, they do not consider the link between censoring and missing observations. To the best of our knowledge, only \cite{aguir} have commented on the link between rank right-censoring and missing data in the context of an extreme value analysis. However, their data analyses included artificial censoring thresholds rather than determining how a block maximum was censored through the missing data assumptions at the series level. 

This work fills a gap in the literature in two ways: (i) by relating the concepts of missing observations within blocks to the right-censoring of block maxima and (ii) by proposing various novel parametric maximum likelihood estimation procedures and studying their performances under a wide variety of simulation settings. In Section 2, relevant notation is defined and the connection between missing observations and censoring is discussed in detail. We propose various estimation procedures using techniques from missing data analysis, survival analysis and computational statistics in Section 3. In addition, we highlight how the assumptions on the type of missingness within blocks can inform the applicability of particular estimation procedures. In Section 4, we conduct a simulation study to compare the various estimation procedures for distributions in the different GEV domains of attraction. These procedures are then used to model the extremal wave surge height return levels at various locations in Atlantic Canada in Section 5. Section 6 contains concluding remarks and a discussion on future research directions. 

\section{Extremes, Missing Data and Censoring}

\subsection{Block Maxima Distribution}

Let $X_1, ..., X_N$ denote a random sample drawn from the cumulative distribution function $F(\cdot)$. Under the block maxima framework, the random sample is split into $k$ blocks, each of size $n$ such that $kn = N$. Let $M_{n}$ denote a generic block maximum, \com{i.e., $M_n$ equals the $i$th block maxima $M_n^{(i)}$ in distribution, where $M_n^{(i)}=\max (X_{(i-1)*n+1},\dots X_{i*n})$ for $i\in \{1,\dots,k\}$}. From extreme value theory, the asymptotic distribution of the normalized maximum is a generalized extreme value (GEV) distribution by the Fisher-Tippett-Gnedenko theorem \citep{fishe,gnedenko1943,mises1936}. Specifically, suppose there exists sequences of constants $\{a_n>0\}$ and $\{b_n\}$ such that for all $z\in \mathbb{R}$:
\begin{equation*}\label{eq:GEVlim}
\lim_{n\rightarrow\infty} P\left(\frac{M_{n}-b_n}{a_n}\leq z \right) = G(z),
\end{equation*}  
for some non-degenerate distribution $G$. Then, $G$ is the generalized extreme value (GEV) distribution given by:
\begin{equation*}
G(z; \mu, \xi, \sigma)=\begin{cases} 
\exp \left[ - \left\{ 1 + \xi \left( \frac{z - \mu}{\sigma} \right) \right\}^{-1/\xi} \right]&\text{ if } 1 + \xi(z-\mu)/\sigma > 0,\xi\neq0, \\
\exp \left\{ - \exp\left(  \frac{z - \mu}{\sigma} \right)\right\}&\text{ if }z\in \mathbb{R}, \xi=0, \\
\end{cases}
\label{gev}
\end{equation*}  
where $-\infty < \mu < \infty$, $\sigma > 0$, and $-\infty < \xi < \infty$. \com{This result also holds for stationary series, under certain conditions on the form of the temporal dependence \citep{leadbetter1974extreme}. The condition on the temporal dependence, called the $D(u_n)$ condition, is satisfied for stationary series for which extreme events are approximately independent if they are sufficiently far apart. Note that the parameters of the limiting distribution for the stationary series are impacted by the degree of temporal dependence in the extremes and can therefore be different than the parameters of the limiting distribution for the corresponding independent and identically distributed series.} Using the formulation of the GEV distribution, $G$, given above, one can form the observed likelihood function and estimate the unknown parameters through maximum likelihood estimation. \com{See \cite{beirlant2004} for an in-depth explanation of extreme value theory, the extension to stationary series, and estimation. }

\subsection{Missing Extremes}

Suppose only $N_{obs}$ random variables in the sample of size $N$ are observed and $N_{miss}$ are missing. Modifying the notation slightly, let $\mathscr{S}$ and $\mathscr{S}'$ denote the disjoint sets of indices of sizes $N_{obs}$ and $N_{miss}$ for the observed and missing random variables, respectively, such that $\mathscr{S}_N=\mathscr{S} \cup \mathscr{S}'=\{1,\dots,N\}$. Let $\{X_{i}:i\in\mathscr{S}\}$ and $\{X_{i}':i\in\mathscr{S}'\}$ denote the corresponding sets of observed and missing random variables. Under the block maxima framework, we denote the maximum order statistics based on the observed and unobserved random samples within block $j$ by $M_{n_j}$ and $M_{n_j'}$, respectively, where $n_j$ and $n_j'$ denote the corresponding fixed set sizes of the observed/unobserved random variables, summing to $n$. Thus, $n_1 + ... + n_k = N_{obs}$, $n'_1 + ... + n'_k = N_{miss}$ and $N_{obs} + N_{miss} = N$. For probabilistic comparisons between the observed block maximum and true block maximum, see Theorem 1 in Appendix A. 
\

There are various mechanisms for the way in which random variables may be missing. They include \emph{missing completely at random} (MCAR), \emph{missing at random} (MAR) and \emph{missing not at random} (MNAR). For data that are MCAR, the probability that an observation is missing is independent of both the observed and unobserved data (i.e. a random Bernoulli draw). For data that are MAR, the probability that an observation is missing is dependent only on the observed data whereas for data that are MNAR, the probability that an observation is missing is dependent on the unobserved data. For further details on the different types of missing data, see \cite{littl}. 

In this work, we consider \com{three} missing data scenarios in the context of estimating the parameters of a GEV distribution under the block maxima framework\com{, defined below. For the following scenarios, consider $I$ to be the indicator of missingness for a random variable $X$, where $I=1$ if $X$ is observed and 0 otherwise.}
\begin{enumerate}
\item[Scenario I:] Each observation in the series has an equal chance of being unobserved and so each block may or may not have a portion of missing observations. This scenario represents a MCAR missingness mechanism\com{, i.e., $P(I=0\mid\{X_{i}:i\in\mathscr{S}_N\})=P(I=0)$.}
\com{\item[Scenario II:] The probability of a random variable in the series being unobserved depends only on the observed random variables, where the probability of missingness depends on the time index. This scenario represents a MAR missingness mechanism, i.e.,\\ $P(I=0\mid\{X_{i}:i\in\mathscr{S}_N\})=P(I=0\mid\{X_{i}:i\in\mathscr{S}\})$.}
\item[Scenario \com{III}:] If a block contains missing observations, then a known number of the higher order statistics of that block are missing \citep{beirl}. As the probability of missing a higher order statistic depends on the unobserved values of the higher order statistics, this scenario corresponds to a MNAR mechanism\com{, i.e., $P(I=0\mid\{X_{i}:i\in\mathscr{S}_N\})$ cannot be simplified to remove $\{X_{i}':i\in\mathscr{S}'\}$ from the conditioning.}
\end{enumerate} 

\begin{figure}
\centering
\subfloat[Scenario I: Missing Completely At Random]{\includegraphics[scale=0.4]{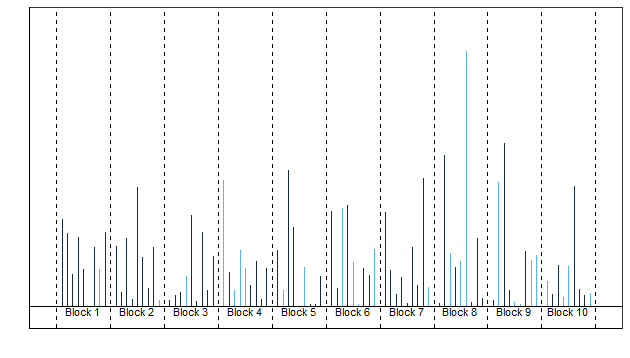}} \quad
\subfloat[Scenario II: Missing At Random (dependent on time index)]{\includegraphics[scale=0.4]{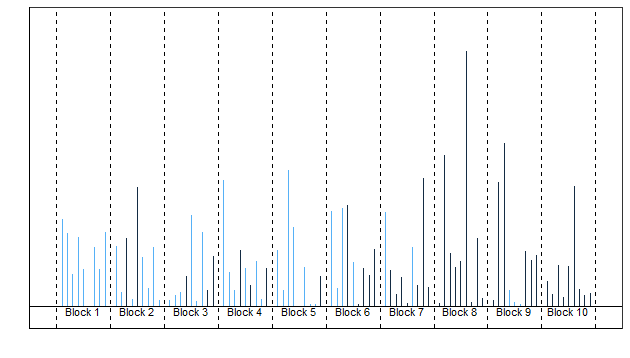}} \quad
\subfloat[Scenario III: Missing Higher Order Statistics]{\includegraphics[scale=0.4]{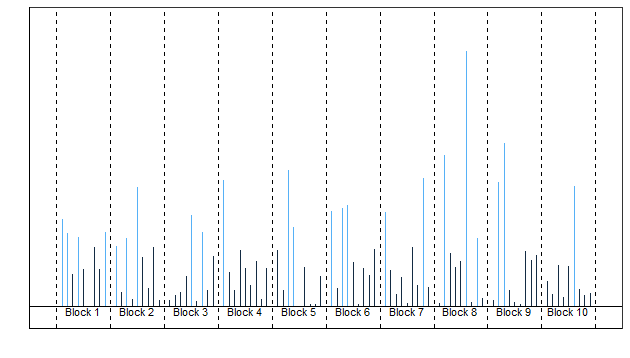}}
\caption{Graphical representation of missing data Scenarios I (Panel (a)), \com{II (Panel (b))} and \com{III} (Panel (\com{c})) in a single block. The observed data are given in black whereas the missing data are given in blue.}
\label{fig1}
\end{figure}

\begin{figure}
\centering
\subfloat[Observed (Black) and Unobserved (Blue) \com{Data}]{\includegraphics[scale=0.5]{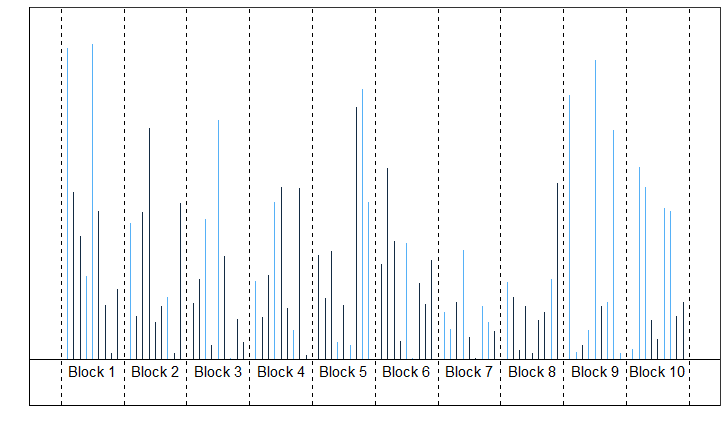}} \\
\subfloat[Observed \com{Data}]{\includegraphics[scale=0.5]{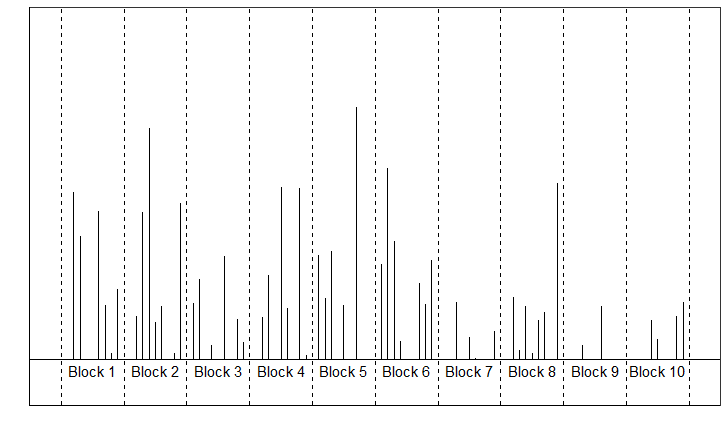}}
\caption{Graphical representation of the effect of missing observations in the context of a block maxima analysis. Panel (a) includes both the observed \com{data} (black) and the unobserved \com{data} (blue) over 10 blocks whereas Panel (b) only includes the observed \com{data} over the same 10 blocks.}
\label{fig2}
\end{figure}

For a visual representation of the differences between missing data scenarios I-\com{III}, see Figure \ref{fig1} panels (a)-\com{(c)}, respectively. \com{In the panel for scenario I, the missing observations are found randomly  throughout the entire series whereas in the panel for scenario II, the proportion of missing observations increases as the time index increases. In the panel for scenario III, approximately 25\% of the largest observations are missing irrespective of the time index.} \com{We note that although all three missingness mechanisms are plausible, MAR or MNAR missingness mechanisms tend to occur more often in real data analyses.} For example, in the context of estimating extreme wave surges measured by tidal buoys, scenario I would indicate that failures of the measuring equipment were completely independent of the intensity of the waves, \com{an example of scenario II would be progressive failure of the measuring equipment over time due to the equipment's prolonged usage}, whereas scenario \com{III} could indicate that a failure of the equipment was attributed to the high intensity waves. If a block has missing observations, in the first \com{or second} scenario, it is typically unknown whether the true block maximum is missing, whereas in the \com{third} scenario, it is known that the true block maximum is missing. However, the \com{third} scenario is not the only possible MNAR scenario and hence, such a missingness mechanism may also mean that the determination of whether the true block maximum is missing is unknown. Figure \ref{fig2} represents an underlying process generating \com{series level data} and how missing observations can yield missing maxima in certain blocks. For example, in blocks 2, 4, 6, and 8, the observed block maxima correspond to the true block maxima, whereas in all other blocks, the true block maxima are unobserved. This missing data feature can severely impact the statistical inferences as the observed maxima in certain blocks are nearly half the size of the true, respective, maxima.       
\

We base our analyses and proposed methods on these scenarios as the motivating example of this research is based on measuring extreme wave surges on the Eastern coast of Canada and many datasets are likely to have missingness mechanisms \com{that resemble one of these cases}. Therefore, it is important that estimation procedures for handling missing data are robust to different missingness mechanisms. This property will be assessed through simulation studies as \com{the occurrence of} our chosen scenarios \com{is plausible when applied to tidal data.} As tidal readings are recorded on an hourly basis, we assume throughout this paper that the total number of measurements (observed/unobserved) per block, $n_j + n_j' = n$ is known in both scenarios.

\subsection{Right-Censoring Equivalency}

The concepts of missing data and censoring are closely related as depending on the missingness mechanism, if a block has missing observations, then it may imply the true block maximum is right-censored. Specifically, under the missing data scenario~\com{III}, as a known number of higher order statistics are missing, then the observed block maximum will correspond to a lower order statistic. Thus, the true block maximum is right-censored by the observed block maximum with probability 1. This equivalency does not hold for all missing data problems under \com{scenarios I and II, as in such cases,} if a block has missing observations, it is not determined with absolute certainty whether the true block maximum will be missing. As will be discussed in the following section, we propose estimators for the probability of whether the true block maximum is missing and show how these estimators can be incorporated into the estimators of the GEV distribution parameters. 

\section{Methodology}

In this section, we propose various procedures to estimate the GEV parameters in the presence of missing data. The motivation for these modelling procedures is drawn from estimation techniques in general likelihood inference, robust estimation and survival analysis.

\subsection{Observed Likelihood}

The simplest technique to estimate the GEV parameters using block maxima, in the presence of missing observations, is to ignore the missing data and maximize the observed data log-likelihood function given by Equation \eqref{obslike} 
\begin{equation}
\ell(m_{n_1}, ..., m_{n_k}; \mu, \xi, \sigma) = \sum_{j=1}^{k} \log\left\{g(m_{n_j}; \mu, \xi, \sigma)\right\}, 
\label{obslike}
\end{equation}
where $m_{n_1}, ..., m_{n_k}$ are the realizations of the $k$ observed block maxima and $g$ is the GEV density function. Although this approach is straightforward, it can lead to underestimation of extreme quantiles as the true maxima within blocks are possibly unobserved. However, as proposed in \cite{beck}, if the missingness level is above a particular threshold, the block maxima with more missing observations can simply be removed from the analysis. 

\subsection{Censored Likelihood Approaches}

Given the equivalency between missing observations within blocks and right-censored maxima, the observed data likelihood will not be efficient as it does not fully account for the position of the block's true maximum relative to the block's observed maximum. We consider two types of censored likelihood approaches: a ``hard'' censored likelihood approach inspired by scenario~\com{III} and a ``soft'' censored likelihood approach inspired by \com{scenarios~I~and~II}.
\
\subsubsection{Hard Censored Likelihood Approach}
In missing data scenario \com{III}, when it is known that the upper order statistics are missing within a particular block, then it is necessarily true that the true block maximum is right-censored by the observed block maximum. In the survival analysis literature, the occurrence of right-censoring within a particular block (i.e. at the series level) by the order statistics is commonly referred to as Type II right-censoring \citep{ander}. Let $(M_{n_j}, \delta_j)$ for $j=1,\dots,k$ denote the pairs of observed maxima and censoring indicator functions, where $\delta_j = 1$ if the $j$th block's observed maximum equals the $j$th block's true maximum and $\delta_j = 0$ otherwise. We refer to this framework as the ``hard censoring approach'' as  $\delta_j$ only takes values in $\{0,1\}$ for all $j=1,\dots,k$. Under this framework, the corresponding likelihood function based on the observed block maxima and censoring indicators is given by Equation \eqref{hrdlike}:   
\begin{equation}
\mathscr{L}(m_{n_1}, ..., m_{n_k}; \mu, \xi, \sigma) = \prod_{j=1}^{k} \left\{g(m_{n_j}; \mu, \xi, \sigma)\right\}^{\delta_j} \left\{\bar{G}(m_{n_j}; \mu, \xi, \sigma)\right\}^{(1 - \delta_j)} ,
\label{hrdlike}
\end{equation}
where $\bar{G}(\cdot)$ is the GEV survival function. 

We note that in missing data scenario \com{III}, the $\delta_j$ are known, for $j=1,\dots,k$. In this case, maximization of \eqref{hrdlike} is justified given that it is known that the upper order statistics within blocks are missing if there is missingness within the block. However, in missing data \com{scenarios I and II}, $\delta_j$ for $j=1,\dots,k$ are not all known, rather, only $\delta_j=1$ is known if block $j$ is fully observed. Therefore, in \com{these scenarios}, maximization of \eqref{hrdlike} with $\delta_j\in \{0,1\}$ for $j=1,\dots,k$ may produce biased estimates of the GEV distribution parameters, as missing observations do not necessarily correspond to missing extremes. 

\subsubsection{Soft Censored Likelihood Approach}

Given that scenarios commonly exist where missing observations within particular blocks may not necessarily indicate that a block's true maximum is right-censored by the block's observed maximum, such as \com{scenarios I and II}, we propose estimation methods for the case where some of the $\delta_j$'s are unknown. More formally, let $\mathscr{D}_u$ and $\mathscr{D}_l$ denote the sets of block maxima indices of sizes $k_u$ and $k_l$, $k_u + k_l = k$, for blocks with missing and no missing data, respectively. Then, we can partition the set of observed maxima given by $\boldsymbol{M}$ into the disjoint subsets $\boldsymbol{M}_u= \{M_{n_j}:j\in \mathscr{D}_u\}$ and $\boldsymbol{M}_l= \{(M_{n_j},\delta_j):j\in \mathscr{D}_l\}$ where $\boldsymbol{M} =\boldsymbol{M}_u\cup\boldsymbol{M}_l$. Note that if the $j$th block is fully observed, then $M_{n_j}\in \boldsymbol{M}_l$ and $\delta_j=1$, otherwise if a block has missing observations then $M_{n_j}\in \boldsymbol{M}_u$ and $\delta_j$ is not observed. 

In this setting, the unknown censoring indicators are regarded as latent unobserved variables. With this view and through parallels to semi-supervised learning, an Expectation-Maximization (EM) algorithm can be used to estimate the GEV model parameters \citep{demps}. The EM algorithm iterates between an Expectation- (E-) step and a Maximization- (M-) step until a convergence criterion is satisfied signalling convergence to an optimum. Let $(\boldsymbol M_u, \boldsymbol \delta_u, \boldsymbol M_l, \boldsymbol \delta_l)$ denote the complete data composed of both observed and unobserved variables where, as before, subscripts $u$ and $l$ denote the unlabeled and labeled observed block maxima, respectively, i.e., $\boldsymbol M_u= \{M_{n_j}:j\in \mathscr{D}_u\}$, $\boldsymbol \delta_u= \{\delta_j:j\in \mathscr{D}_u\}$, $\boldsymbol M_l= \{M_{n_j}:j\in \mathscr{D}_l\}$, and $\boldsymbol \delta_l=\boldsymbol{1}= \{\delta_j:j\in \mathscr{D}_l\}$. Hence, the unlabeled right-censoring indicators $\boldsymbol \delta_u$ are regarded as latent data components. For simplicity of exposition, we denote the GEV parameter vector $(\mu, \xi, \sigma)$ by $\Theta$. The complete data log-likelihood function is given by:
$$\ell_c(\Theta; \boldsymbol M_u, \boldsymbol \delta_u, \boldsymbol M_l, \boldsymbol \delta_l) = \sum_{j=1}^{k} \left[ \delta_{j} \log g(m_{n_j}; \Theta) + (1 - \delta_j) \log \left\{\bar{G}(m_{n_j}; \Theta)\right\} \right],$$
which can be re-expressed in terms of the unlabeled and labeled block maxima, \textit{viz.}
 $$\ell_c(\Theta; \boldsymbol M_u, \boldsymbol \delta_u, \boldsymbol M_l, \boldsymbol \delta_l) = \sum_{j\in \mathscr{D}_u} \left[ \delta_j \log g(m_{n_j}; \Theta) + (1 - \delta_j) \log \left\{\bar{G}(m_{n_j}; \Theta)\right\} \right] + \sum_{j\in \mathscr{D}_l} \log g(m_{n_j}; \Theta).$$
Then, the corresponding minorizing function in the E-step at iteration $t+1$ is given by:
 \begin{align*} 
 Q(\Theta \mid \Theta_t) &= \sum_{j\in \mathscr{D}_u} \big[ \mathbb{E}(\delta_j\mid\boldsymbol M_u, \boldsymbol M_l, \boldsymbol \delta_l, \Theta_t) \log g(m_{n_j}; \Theta)\\ & \qquad\qquad+ \{1 - \mathbb{E}(\delta_{n_j}\mid\boldsymbol M_u, \boldsymbol M_l, \boldsymbol \delta_l, \Theta_t)\} \log \left\{\bar{G}(m_{n_j}; \Theta)\right\} \big] 
  + \sum_{j\in \mathscr{D}_l} \log g(m_{n_j}; \Theta),
 \end{align*}
where the conditional expectation of $\delta_j$ for $j\in \mathscr{D}_u$ can be simplified to:
 \begin{equation} 
 \mathbb{E}(\delta_j\mid\boldsymbol{M}, \Theta_t) = \mathbb{P}(\delta_j = 1 \mid M_{n_j}, \Theta_t)  = \mathbb{P}(M_{n_j'} \leq M_{n_j}\mid M_{n_j}, \Theta_t).
 \label{eqn:exp}
 \end{equation} 
 The corresponding M-step is:
 \begin{equation*} 
\hat\Theta_{t+1}=\argmax_\Theta Q(\Theta \mid \Theta_t).
 \end{equation*}

Let $\hat{\delta}_j$ denote the expectation in \eqref{eqn:exp}. Knowledge of the missingness mechanism dictates how this expectation should be evaluated. If this expectation depends on the GEV model parameters, the EM algorithm iterates until satisfying a lack-of-progress stopping criterion based on the updated GEV parameter estimates. Otherwise, the EM algorithm is complete in a single iteration. Given that $\hat\delta_j \in [0, 1]$, the following EM algorithm based methods will be referred to as ``soft-censoring'' approaches.
\

Suppose the missingness mechanism is MCAR. Within block $j$, unconditionally on the observed maximum within that block, $\hat\delta_j$ is the proportion of non-missing observations within the block, i.e., $\hat{\delta}_j = \frac{n_j}{n_j + n_j'}$ (see Theorem 2 (ii) in Appendix A for additional details). Thus, the complete log-likelihood contribution for block $j\in \mathscr{D}_u$ is the weighted average of the GEV density and survival functions where the weights are determined by the number of non-missing observations within each block. Again, these estimates for $\delta_j$ are unconditional and do not utilize any information from the random variables within each of the blocks. This estimation procedure will be referred to as the ``soft-censoring unconditional'' approach.
\

For missingness frameworks that are not MCAR, we suggest two possible ways of calculating $\hat\delta_j$. First, we note that although Equation \eqref{eqn:exp} expresses the conditional expectation in terms of the observed maxima, the expectation could instead be evaluated by conditioning on all observed data at the series level, i.e., $\{X_{i}:i\in\mathscr{S}\}$. Specifically, the expectation of $\delta_j$, conditionally on all the observed series level data, can be estimated using the empirical cumulative distribution function (CDF) of all observations across all blocks. That is, the conditional expectation of $\delta_j$ is given as follows:
$$\hat{\delta}_j = \mathbb{E}(\delta_j\mid X_{i}:i\in\mathscr{S}) = \left\{ \frac{1}{N_{obs}} \sum_{i\in\mathscr{S}} \mathbb{I}(X_i \leq m_{n_j}) \right\}^{n_j'}= \hat{F}_{N_{obs}}(m_{n_j})^{n_j'},$$
where $\mathbb{I}(\cdot)$ is an indicator function (see Theorem 2 (i) in Appendix A for additional details). As the estimated censoring indicators are determined conditionally using all available data, this estimation procedure will be referred to as the ``soft-censoring conditional'' approach. 

The second approach, and our final proposed estimation procedure, estimates the expectation of $\delta_j$ conditionally on the observed maxima in terms of the GEV CDF, i.e.,
 $$\hat\delta_j=\mathbb{P}(\delta_j = 1 \mid \boldsymbol M_u, \boldsymbol M_l, \boldsymbol \delta_l, \Theta_t) = \mathbb{P}\left\{\max(M_{n_j}, M_{n_j'}) \leq m_{n_j} \mid \Theta_t\right\} = G(m_{n_j}; \Theta_t).$$
From the three proposed estimates of $\hat \delta_j$, the latter is the only estimate which yields an EM algorithm with more than one iteration. Therefore, the term ``EM'' approach will be used to refer to this estimation method even though the previous three estimates were also based on the EM algorithm framework. \

 \textbf{Remark:} Unlike in most uses of the EM algorithm, the so-called ``observed data'' log-likelihood function, i.e., the likelihood using the GEV for the true block maxima, is not easily calculated as the true block maxima may not be fully observed and it is unclear how the blocks with missing observations should contribute to the corresponding likelihood. Thus, we use a lack-of-progress stopping criterion based on the sequential change in the parameter estimates of the GEV distribution: $|\Theta_{t+1} -\Theta_t|<\epsilon$ for some tolerance level $\epsilon > 0$.   \
 
\subsection{\com{Bootstrapping for Observed/Censored Likelihood Approaches}}

\com{The observed and soft/hard censored likelihood approaches discussed above can produce parameter estimates for the underlying GEV distribution using the observed block maxima however a measure of the variability of the parameter estimators is generally required to form confidence intervals. Under the appropriate regularity conditions, the observed Fisher information can be used to estimate the standard errors of the observed likelihood parameter estimators. Using the soft-censoring likelihood approach, as the procedure is based only on the observed maxima and the number of observations within their blocks, the nonparametric bootstrapping procedure of \cite{Efron} can be easily applied. This procedure is based on resampling the observed block maxima with replacement to obtain a bootstrap sample and computing the corresponding soft-censoring likelihood parameter estimates for each such sample. Under the hard-censoring likelihood approach, when the values of $\delta$ are assumed to be either $0$ or $1$, the right-censoring nonparametric bootstrapping procedure of Efron can also be easily applied to obtain parameter standard errors. This procedure is based on resampling the observed maxima and censoring indicator pairs with replacement to obtain a bootstrap sample and then estimating the unknown parameter estimates for each such sample \citep{Efron}. When all the block maxima are fully observed (i.e. $\delta_j = 1$ for $j = 1, ..., |\mathcal{D}_u \cup \mathcal{D}_l|$), then the right-censoring bootstrapping procedure reduces to the standard nonparametric bootstrapping procedure for fully observed i.i.d. data.}

\section{Simulations}

We compared the performance of the methodologies discussed in Section 3 under a variety of different simulation settings. We simulated random samples from either a t-distribution (df = 5), an exponential distribution (rate = 1), or \com{a beta distribution (scales = 2, 5)} corresponding to the maximum domain of attraction of the Fr\'echet, Gumbel, and \com{Weibull} distributions, respectively \citep{beirlant2004}. Under the scenario I missingness assumption, the missing observations were randomly selected within blocks; \com{under scenario II the probability of missingness depended on the time index becoming less likely as time progressed}; and under scenario \com{III} missingness assumption, the upper extremes were removed from the blocks based on the expected proportion of missing observations within the blocks. \com{Due to the missingness of the scenario II relating to the time index rather than to blocks, the settings considered per distribution for this scenario were different than for the remaining scenarios. For scenarios I and III,} we considered a fixed number of observations per block ($n = 100$) but varied the number of blocks (blocks $=25, 50,$ and $100$), the proportion of blocks with missing data (pbm $=0.20, 0.50$ and $0.80$) and the expected proportion of missing observations within the blocks (pm $=0.05, 0.20$, $0.35$). \com{For scenario II, we considered different numbers of observations per block ($n = 50,100$) and again varied the number of blocks (blocks $=25, 50,$ and $100$), then varied the average proportion of missingness for the series data (apm $=0.05, 0.15$ and $0.25$).} We replicated the data generating procedure until $1000$ sets of parameter estimates were obtained for all optimization procedures with successful numerical convergence. \com{We compared the various estimation procedures via the return level estimation performance. The average 50-year return level estimates and standard errors under missingness scenarios I, II, III for the exponential distribution are reported in Tables \ref{simtable1}-\ref{simtable3}. The tabulated results from the t and beta distribution simulations can be found in Tables 1-6 in the Supplementary Materials.}
\

\com{From the simulation results, the performance of the various methods appears more affected by the proportion of missingness and the missingness mechanism rather than the data generating distribution. In addition,} for both scenarios I and \com{III}, the performance of the various methods was more affected by the proportion of missingness
within blocks rather than the proportion of blocks with missing observations. This result appears to support the often-used rule of thumb where blocks with more than 10-15\% of the observations missing are dropped from the analysis, e.g., in \cite{beck}. As previously discussed, the hard censoring approach assumes that if a block has missing observations, then the corresponding block maximum is necessarily right-censored. As this is not generally the case in \com{scenarios I and II, the hard censoring approach had by far the worst performance yielding the extremely large return levels and standard errors. In terms of average absolute deviation from the true return level, the EM and soft unconditional censoring approaches had the next poorest performances in these scenarios, where the EM approach had inflated standard errors in settings with small numbers of blocks (i.e., numbers of blocks = 25).} The soft censored conditional likelihood approach generally outperformed the standard observed likelihood approach in terms of accuracy, \com{but had larger standard errors. Larger standard errors do not indicate a deterioration in performance as in this case it is a reflection on the uncertainty as to whether an observed maxima is a true maxima and therefore appears to be a more realistic reflection of the overall uncertainty.} \comR{The increased standard errors are also a reflection of the larger return levels estimated on average.} In contrast, the performances of the hard censoring and EM approaches were the best in scenario \com{III in terms of average absolute deviation in return levels. However, it is clear from the simulation results that in all scenarios, the hard censoring approach is extremely unstable for small sample sizes (i.e., small numbers of blocks) as opposed to the EM approach which has slightly more instability than other approaches in scenarios I and II and tends to have similar or much smaller variation than the observed approach in scenario III.} Interestingly, the conditional soft censoring approach was comparable to the hard censoring and EM algorithm approaches in some scenario \com{III} cases and still outperformed the observed likelihood estimation approach. The soft unconditional censoring approach did reasonably well in several simulation settings but was not a best performer. Given the mixed results, it is apparent from this simulation study that an estimation approach should be selected for each dataset based on knowledge of the missingness mechanism. 

\begin{figure}[h!]
\centering
\includegraphics[width=0.32\textwidth]{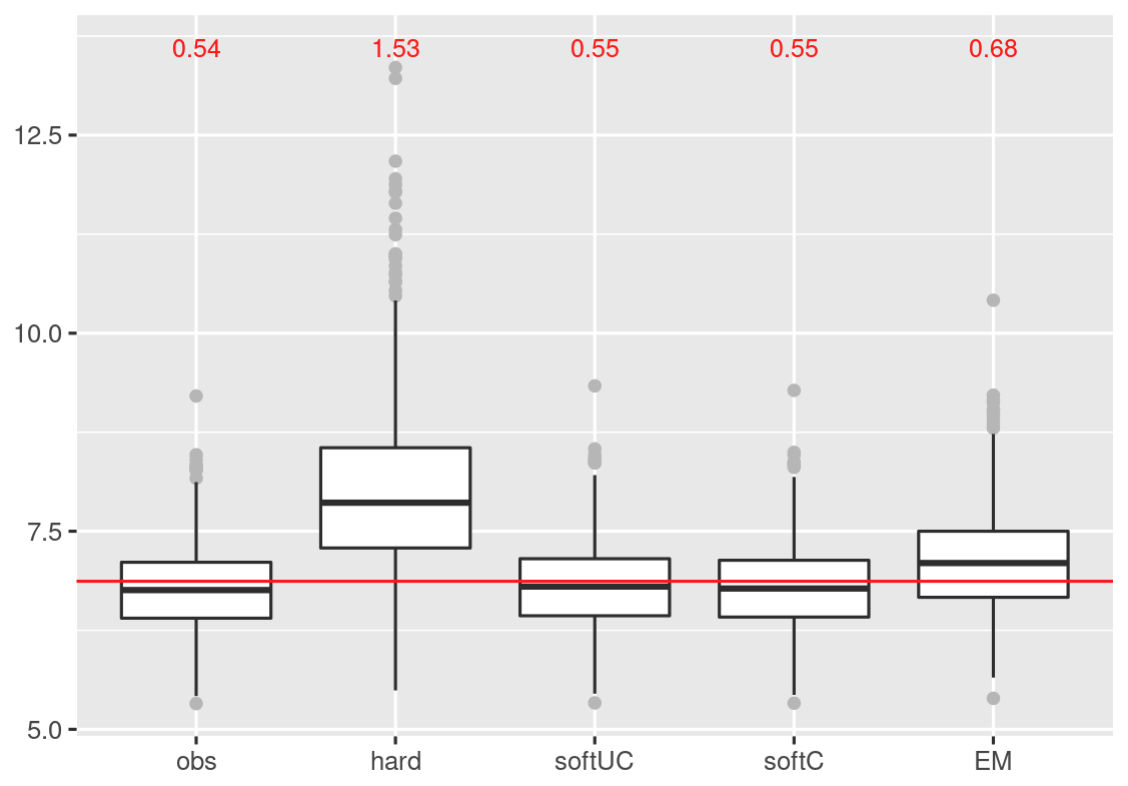}
\includegraphics[width=0.32\textwidth]{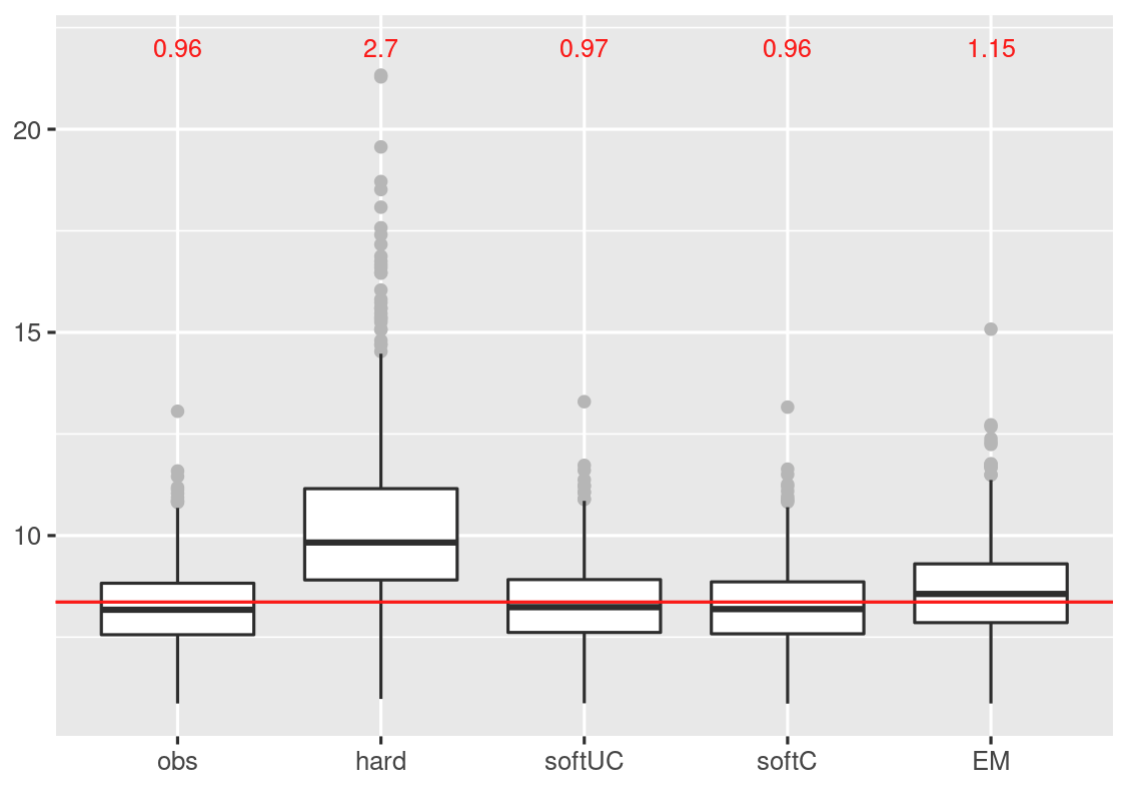}
\includegraphics[width=0.32\textwidth]{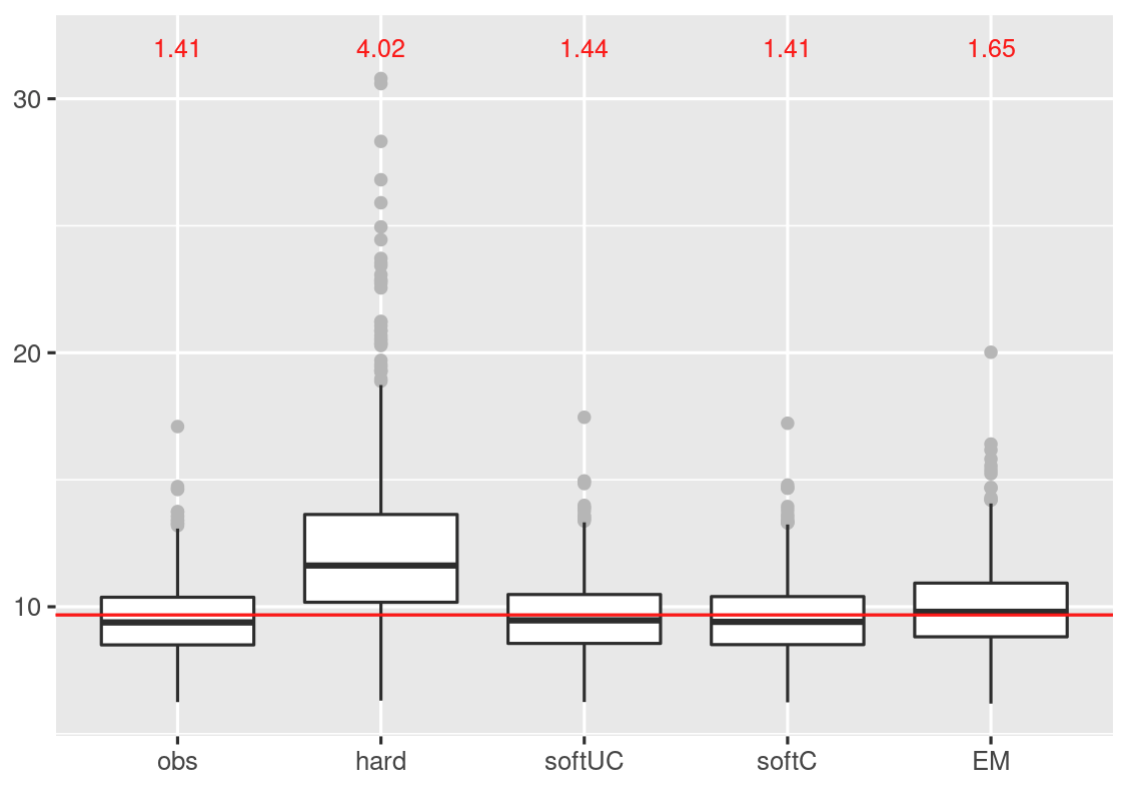}
\caption{\comR{Left to Right:} \com{Comparison of 20-, 50-, and 100-year return levels for a single simulation of scenario I from the t distribution (df = 5) for the five estimation methods over 1000 replications. The simulation setting displayed consists of 100 blocks, where 20\% of blocks had missing values with an average of 5\% missingness. True return levels are plotted as \comR{red horizontal lines and root mean square error for each estimator is in red}.}}
\label{fig3} 
\end{figure}

\begin{figure}[h!]
\centering
\subfloat[Beta distribution (scales = 2,5) - \comR{Left to Right: Missing Completely at Random, Missing at Random, Missing Not at Random}]{\includegraphics[width=0.32\textwidth]{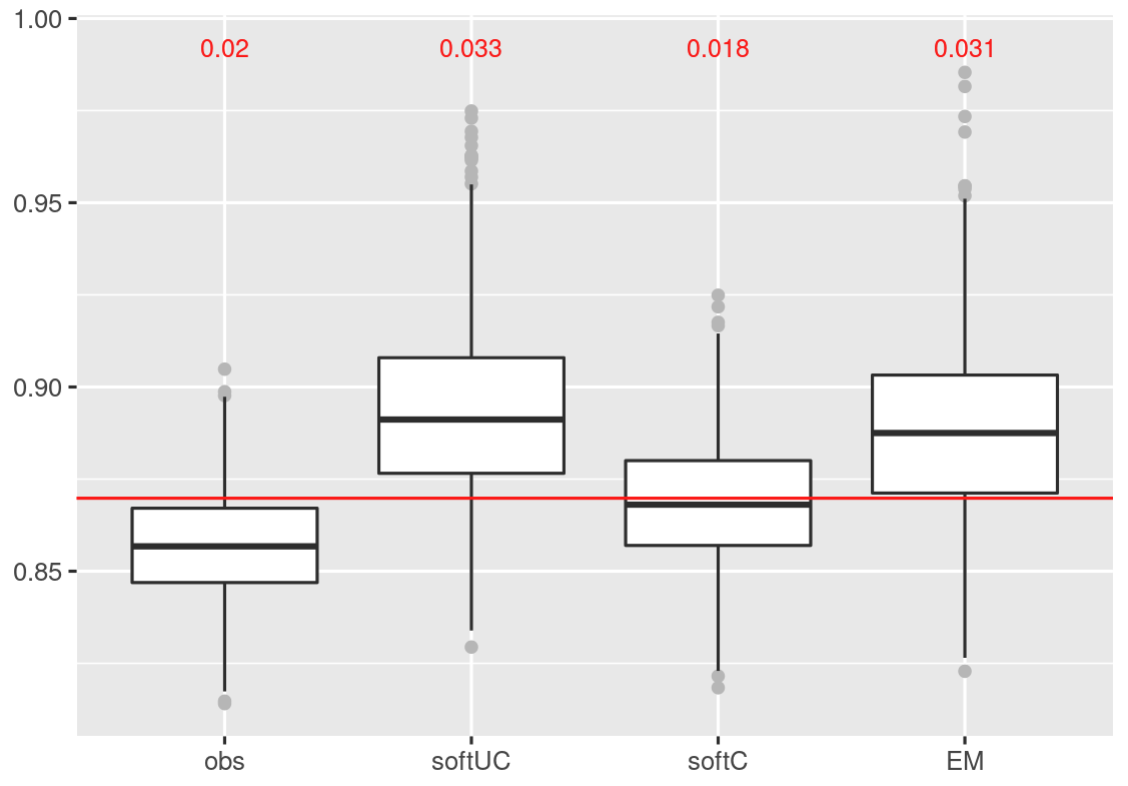}
\includegraphics[width=0.32\textwidth]{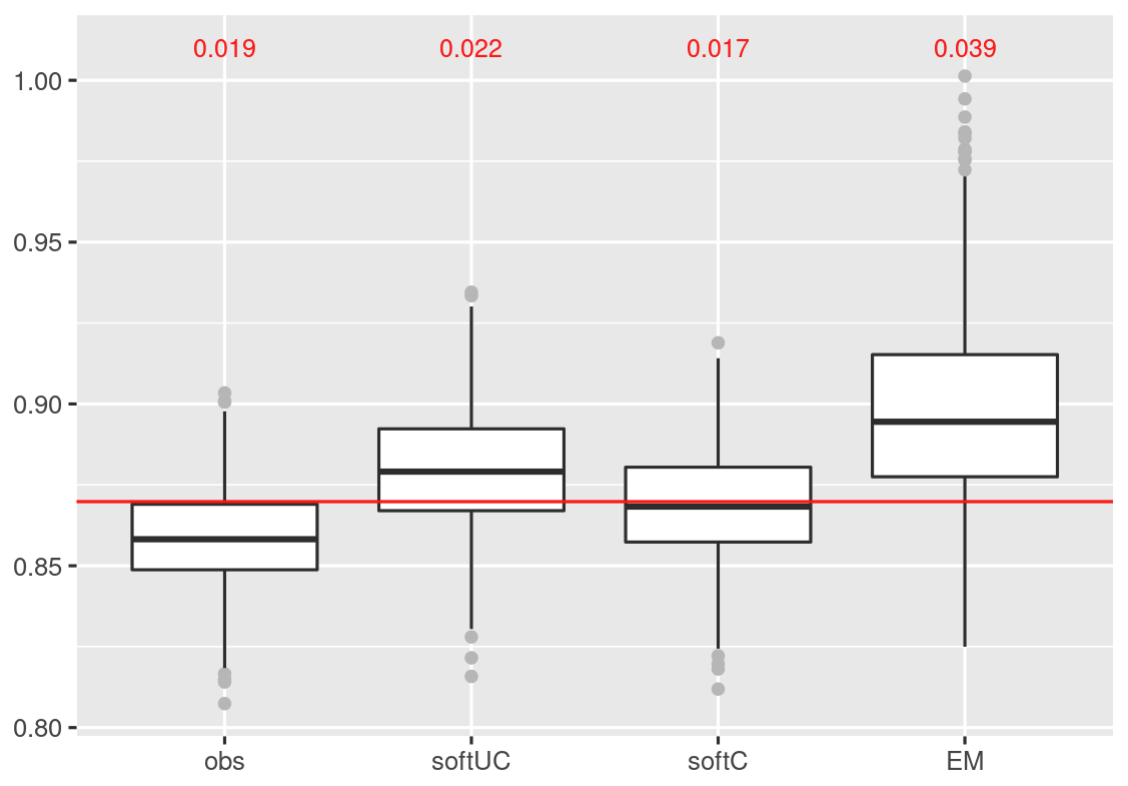}
\includegraphics[width=0.32\textwidth]{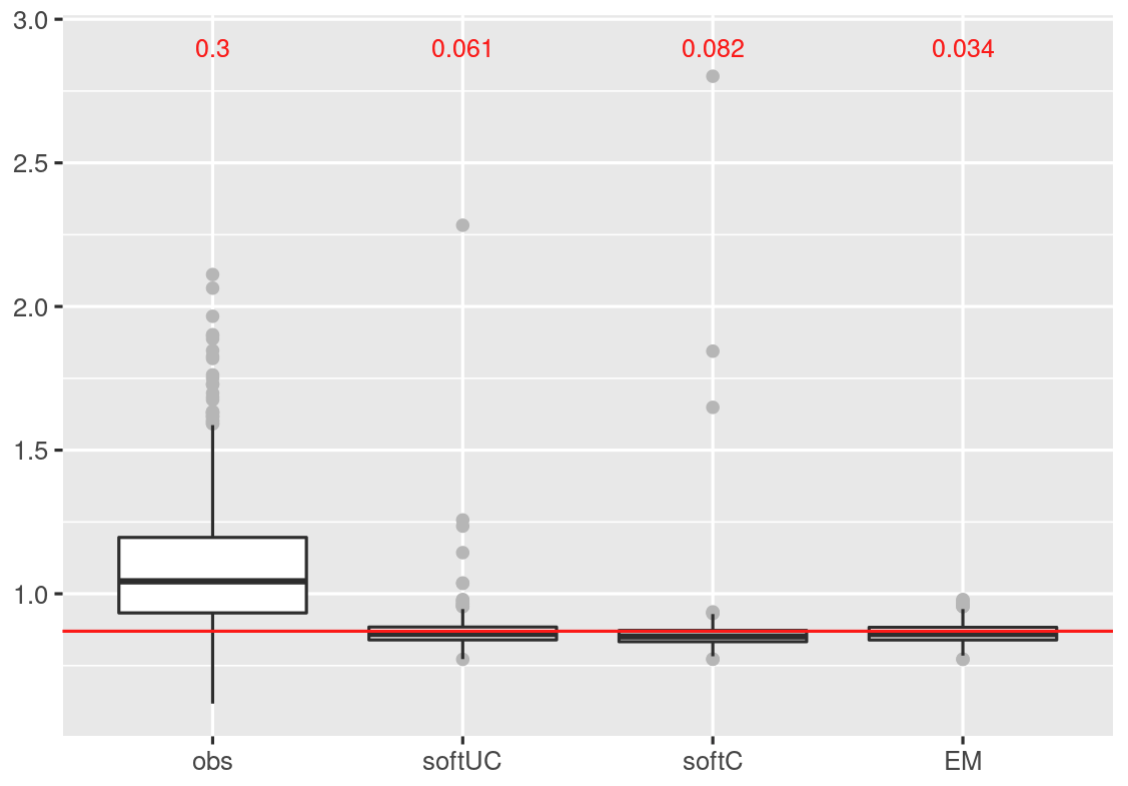}} \\
\subfloat[Exponential distribution (rate = 1) - \comR{Left to Right: Missing Completely at Random, Missing at Random, Missing Not at Random}]{\includegraphics[width=0.32\textwidth]{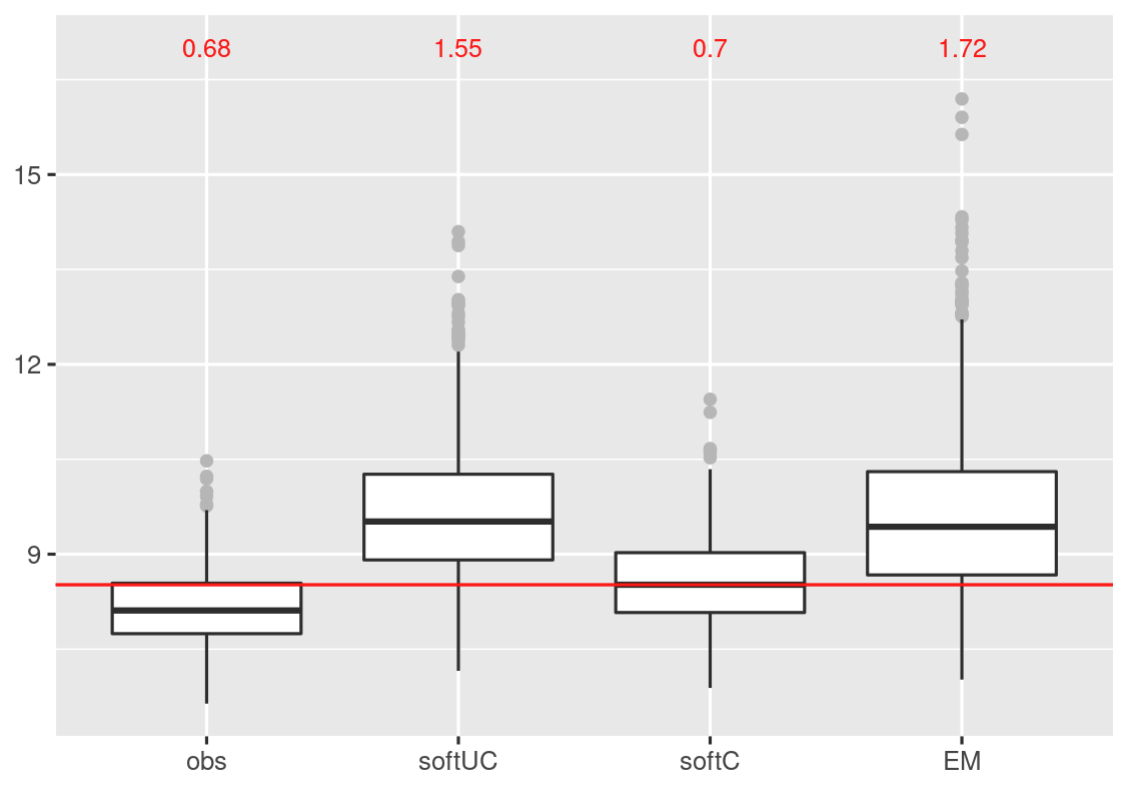}
\includegraphics[width=0.32\textwidth]{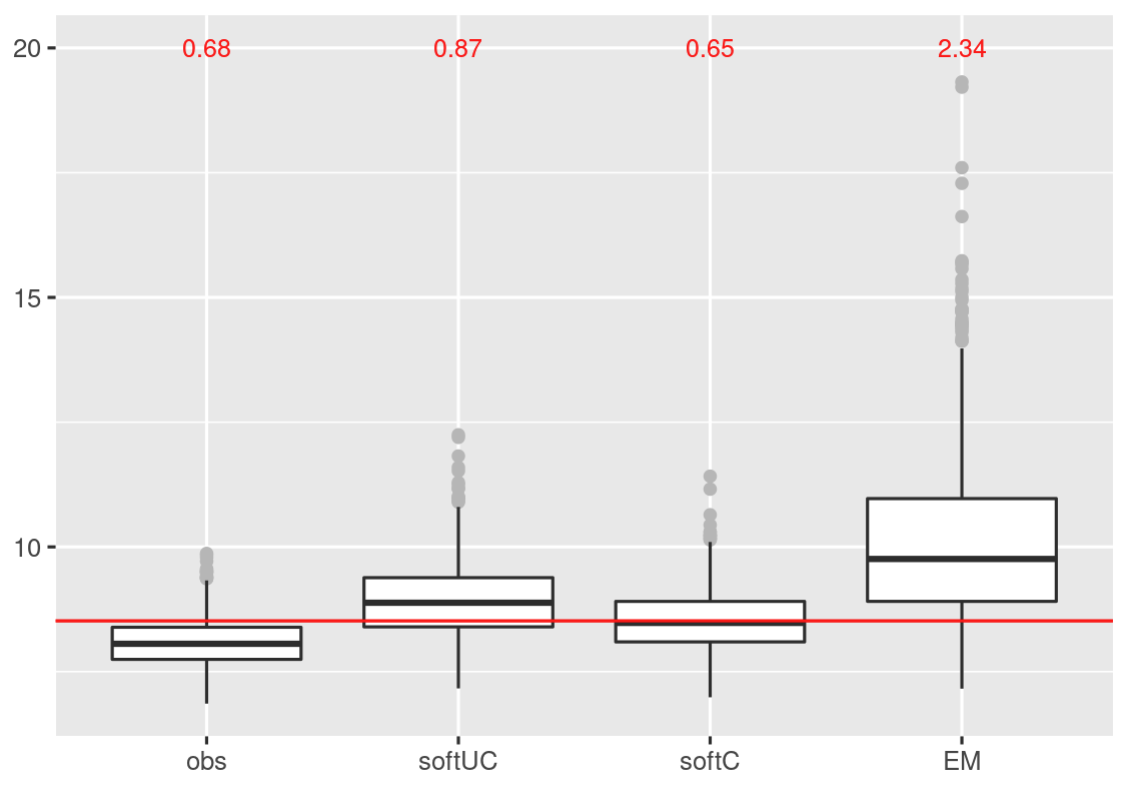}
\includegraphics[width=0.32\textwidth]{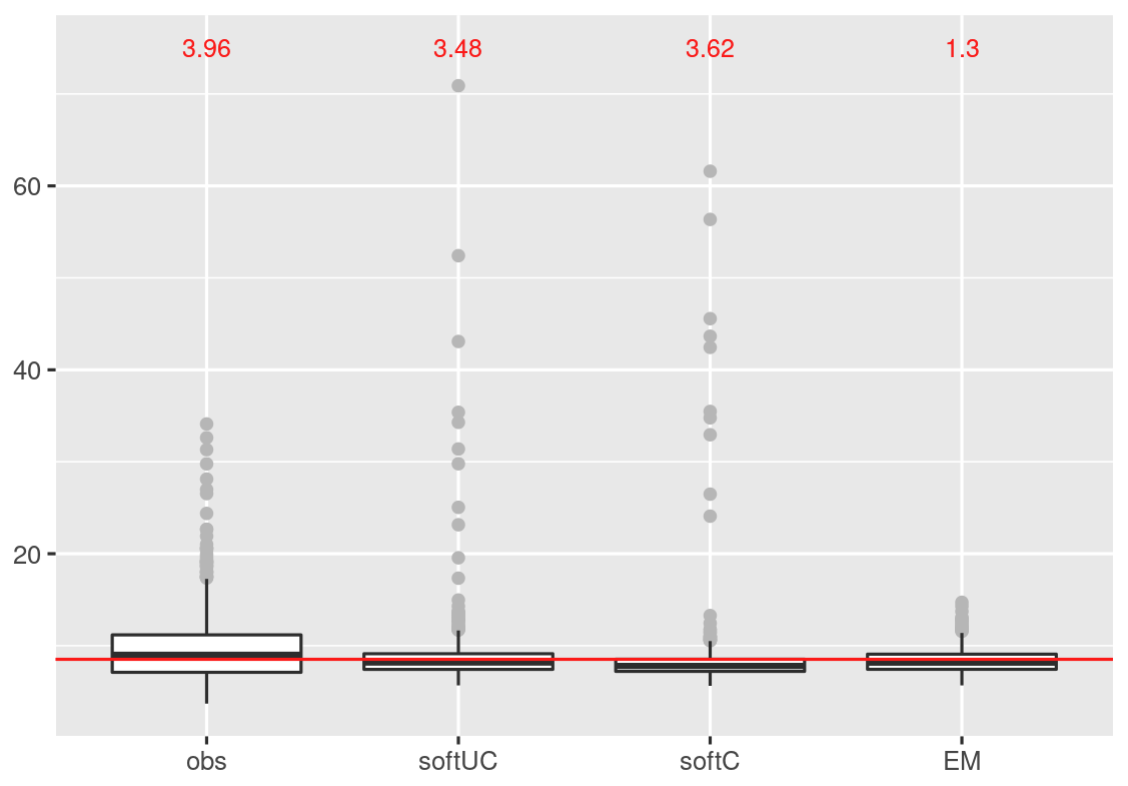}}\\
\subfloat[t distribution (df = 5) - \comR{Left to Right: Missing Completely at Random, Missing at Random, Missing Not at Random}]{\includegraphics[width=0.32\textwidth]{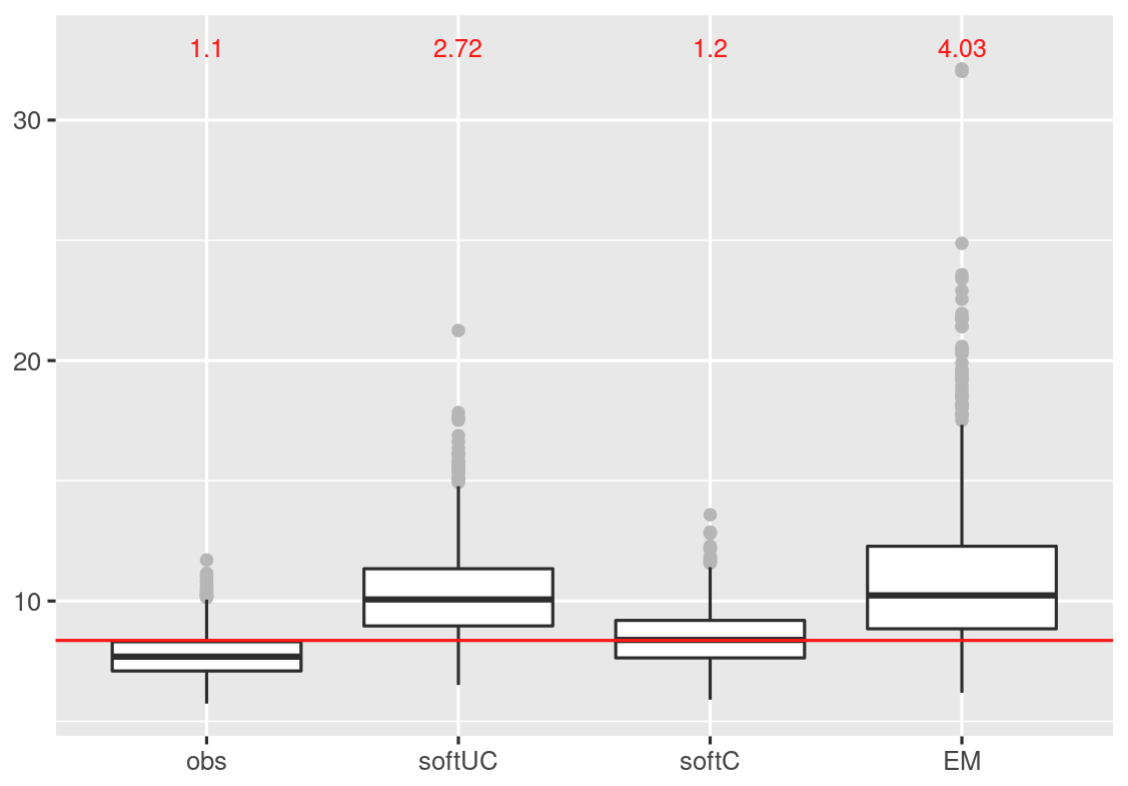}
\includegraphics[width=0.32\textwidth]{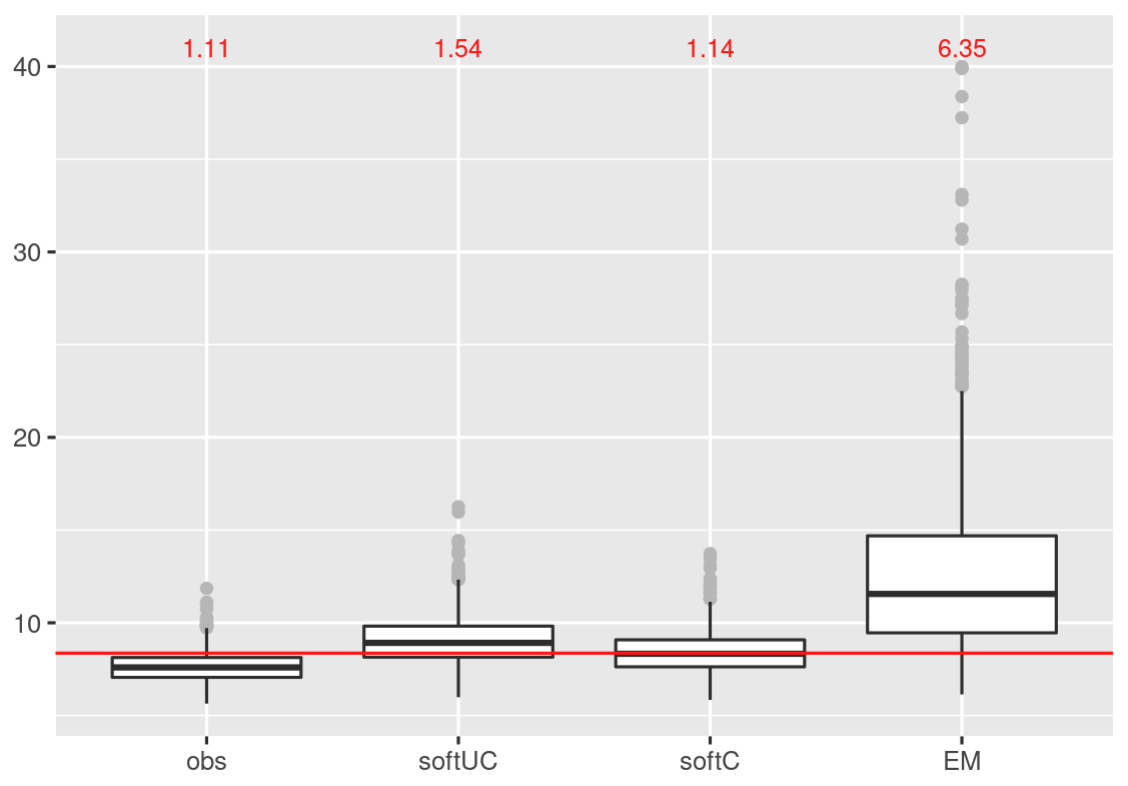}
\includegraphics[width=0.32\textwidth]{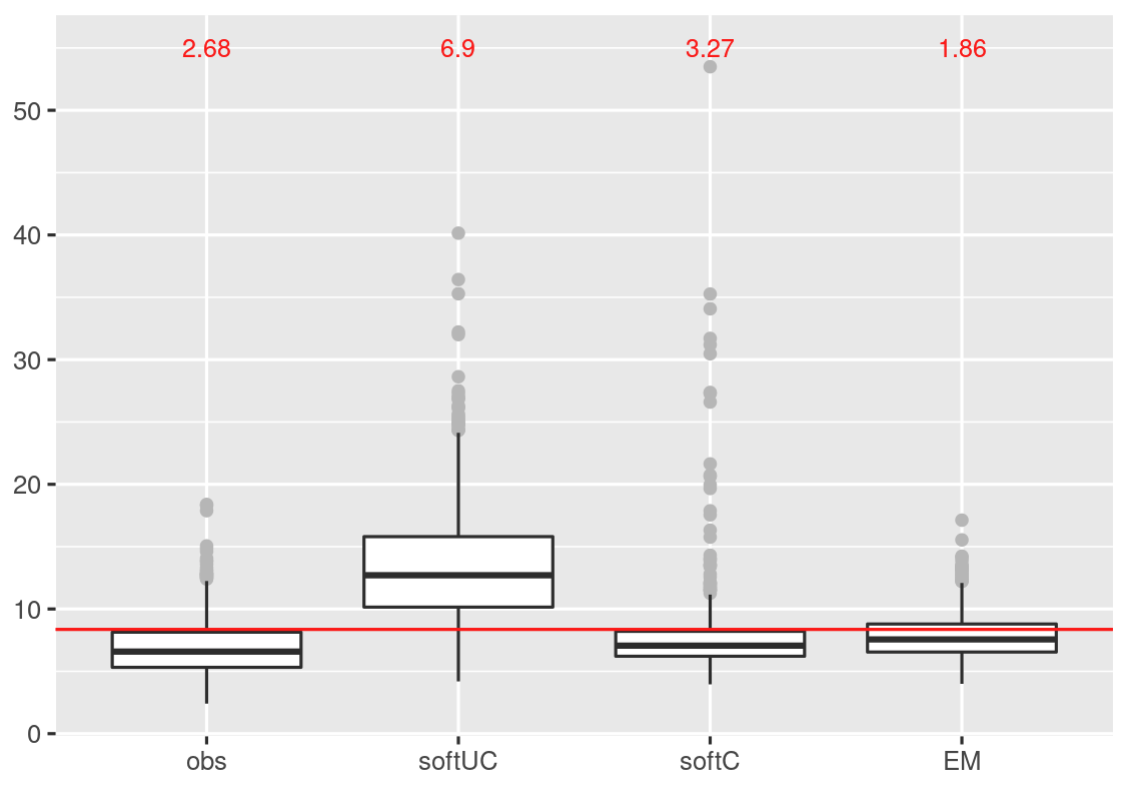}}
\caption{\com{Comparison of 50-year return levels for each missingness scenario over 1000 replications for all estimation methods except hard censoring. Settings for scenarios I and III are 100 blocks with 80\% of blocks having missing data, and 35\% missingness within such blocks. For scenario II, the number of blocks and number of observations within blocks is 100 and average percentage of missingness is 25\%. True return levels are plotted as \comR{red horizontal lines and root mean square error for each estimator is in red}.} }
\label{fig4}
\end{figure}

\com{Select simulation settings are illustrated in Figures \ref{fig3} and \ref{fig4}. Figure \ref{fig3} provides an example of a visual comparison of the estimated 20-, 50-, and 100-year return levels compared to the true return levels (plotted by \comR{red lines}) for each of the estimation methods over 1000 replications of a single case from scenario I with data generated from the t distribution (df = 5). The specific setting had 100 blocks where 20\% of these blocks had missing values each with an average of 5\% missingness. It is apparent from the boxplot in Figure \ref{fig3} that the performance changes as expected with increasing return periods yielding increasing variation in estimation. However, the relative performance between methods remains similar across return periods. It is also clear from this figure and from the tabulated simulation results that the hard censoring return level estimates are much less stable. For this reason, these estimates have been removed from the Figure \ref{fig4} simulation boxplots to ensure a more reasonable scale. Figure \ref{fig4} illustrates a comparison in performance of methods across the different scenarios \comR{for the three data generating distributions}. These visuals support the findings from the tables that the soft-censoring \comR{conditional} approach appears to have the best overall performance for MCAR and MAR scenarios \comR{in terms of bias and is similar to the observed likelihood estimation approach in terms of root mean square error, a reflection on differences in uncertainty as mentioned previously. The ``EM'' method appears to be the superior approach overall} for the MNAR scenario.}

\section{Extreme Wave Height Modelling}

A motivating example for the development of extreme value estimation procedures in the presence of missing observations is the analysis of extreme wave surges measured by coastal buoys or tidal gauges. Such analyses are used to facilitate future predictions of coastal regions that are prone to flooding \citep{rychl, muis, ross, towe, beck}. Buoys are subject to extreme wind speeds and destructive forces by crashing waves which consequently lead to the tidal gauge huts being subject to power loss and flooding. The loss of power and flooding of the measurement apparati can result in gaps in the observed data series. Thus, with the presence of missing observations, when wave height maxima are obtained on a yearly basis, the observed maximum wave height may not correspond to the true maximum wave height during that year \citep{ryden}. 
\

\begin{figure}
\centering
\includegraphics[scale=0.6]{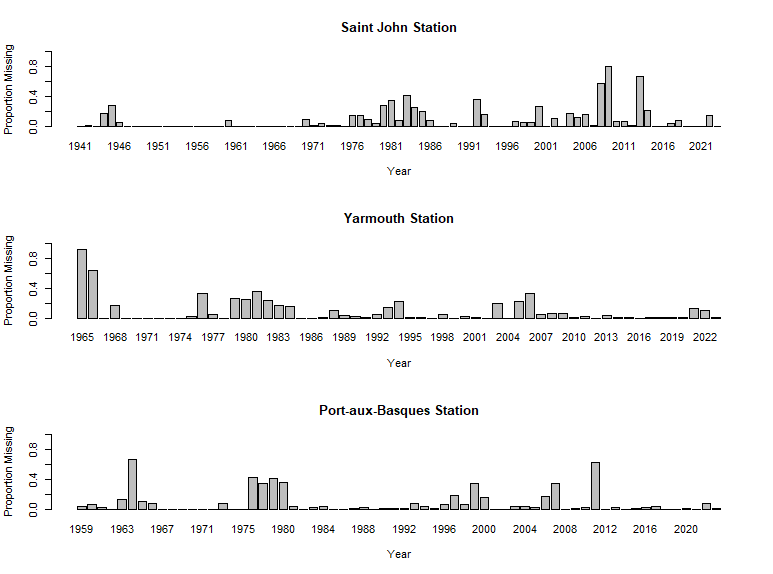}
\caption{\com{A barplot of the proportion of missing observations measured by year for the Saint John (1941-2023), Yarmouth (1965-2023) and Port-aux-Basques (1959-2023) stations.}}
\label{fig5}
\end{figure}

The estimation of extreme wave surge return levels is crucial for determining the risk of flooding in specific coastal regions \citep{ryden,rychl,muis, ross, towe}. Hourly water levels are measured at several locations along the Eastern coast of Canada. These measurements were collected for varying lengths of time and are freely accessible (\url{https://www.tides.gc.ca/}). Our analysis incorporates historical data from three locations on the Atlantic coast of Canada: Saint John (ref \#: 65), Yarmouth (ref \#: 365) and Port-Aux-Basques (ref \#: 665), with corresponding time periods of  1941-2023 (83 years), 1965-2023 (59 years) and 1959-2023 (64 years), respectively. The number of years with missing observations for each location were 60, 45 and 56 with corresponding mean proportions of missing data in years with missing observations of 12.11\%, 12.02\% and 10.67\%, respectively. \com{A bar plot of the proportion of missing observations by year for each of the stations is given in Figure \ref{fig5}.}

\begin{figure}
\centering
\includegraphics[scale=0.5]{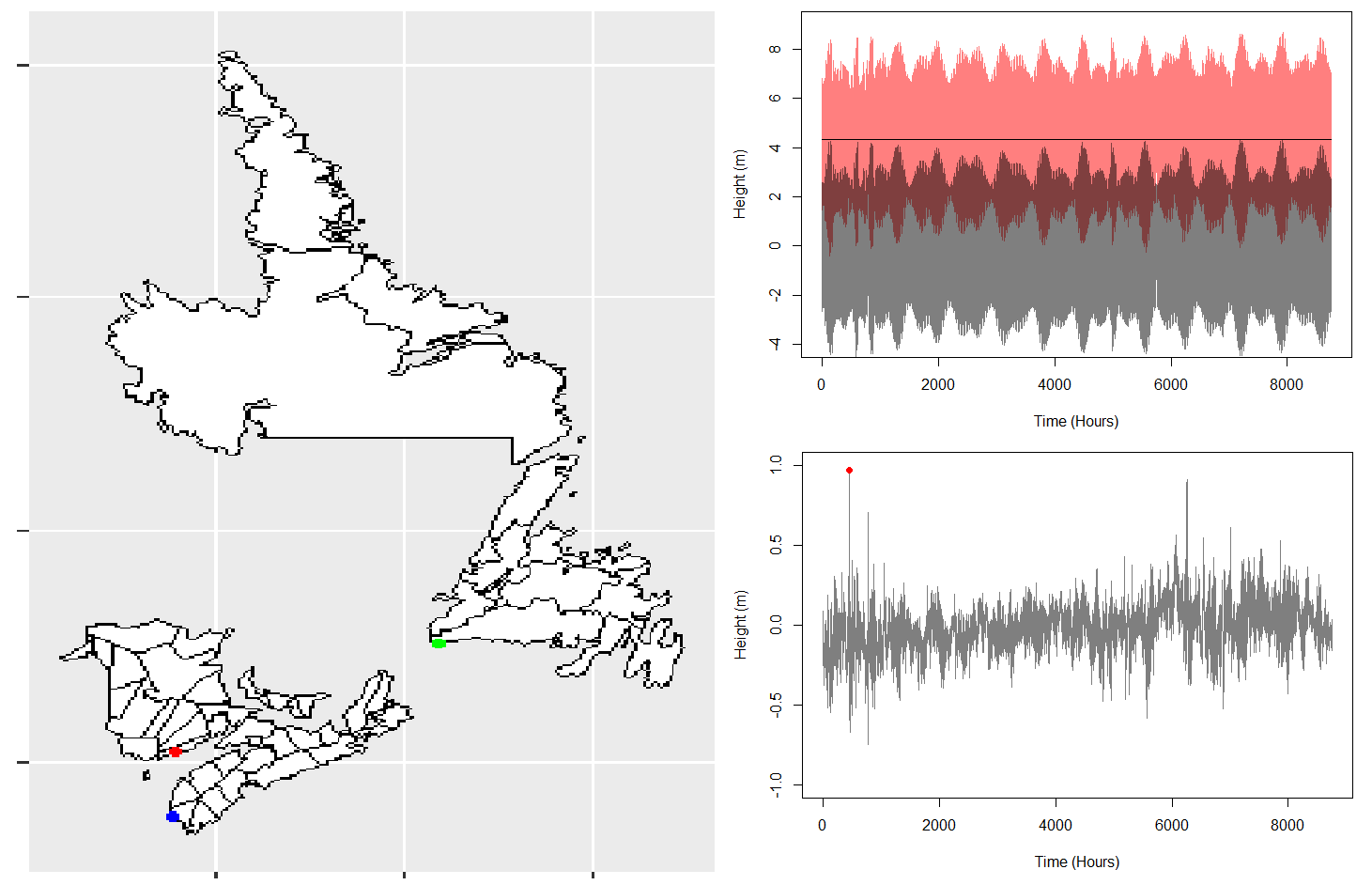}
\caption{\com{Left Panel: Geographical locations of the Saint John (red), Yarmouth (blue) and Port-Aux-Basques (green) water level measurement stations in Atlantic Canada. Top-right Panel: An application of the deterministic tidal model applied to the Saint John station data (levels measured in metres) measured over a one year period. The raw wave data are given in red, the mean sea level is given by the solid black line and the sinusoidal fitted tidal components are given in grey. Bottom-right Panel: The random surge levels after the removal of the deterministic components. The red dot in the surge level series corresponds to the maximum wave level for that year.}}
\label{fig6}
\end{figure}

Hourly water levels can be decomposed into a deterministic tidal process and a stochastic process corresponding to surges. As in \cite{beck}, the deterministic tidal component at time $t$ is modelled using:
$$T(t) = M(t) + \sum_{n=1}^{N} A_n \cos\left[ \pi \left\{\omega_n t - \psi_n\right\}/180\right],$$
where $M(t)$ corresponds to the time-varying mean sea-level and the summation expression corresponds to the regular high/low tide pattern in the waves. \com{In general, the various components of the tidal component formula are estimated through a process based on linear regression where the $M(t)$ function is nonparametrically estimated through a LOESS type smoothing function.} The model was fit using the TideHarmonics $\mathtt{R}$ package \citep{steph} to each location separately. After determining an estimate of $T(t)$ at each recorded time $t$, the estimated tidal component is subtracted from the observed wave heights to obtain an estimate of the stochastic wave surge series. \com{An illustration of the estimated mean sea level, tidal component and resulting stochastic surge level for the Saint John station is given in Figure \ref{fig6}}. As the missingness mechanism is unknown, we applied Little's test to each station to test whether the missingness mechanism was MCAR \citep{jamshidian2010tests, littl}. In all cases, there was no evidence that the missing data mechanisms were MCAR. We applied the various estimation procedures discussed above to the estimated wave surges to estimate the GEV model parameters and the 20-, 50- and 100-year return levels. We calculated approximate estimator standard errors \com{and 95\% empirical confidence intervals} for each method using the nonparametric bootstrap procedure of Efron by either resampling the observed maxima with replacement or resampling the pairs of observed maxima and corresponding $\delta$ estimates \citep{Efron}. The results are summarized in Table \ref{locations}. 
\

Most methods predict similar surge height return levels, with the exception of the hard censoring method. This result is expected as the hard censoring model assumes that blocks with missing observations have true block maxima greater than what was observed. Interestingly, other than at the Saint John location, the EM approach estimated return levels that were in between the hard censoring and the soft-censoring methods. Indeed, at this location, both the EM and conditional soft censoring approaches predict a 20-year return level almost 1m higher than the predicted return level from the observed likelihood method. These results suggest that waves can potentially reach large heights than typically predicted, posing a significant threat to coastal regions. Furthermore, the increased standard errors of the proposed methods, particularly for the hard censoring, conditional soft censoring and EM estimators, reflects the uncertainty associated with some of the observed block maxima.  

\section{Discussion}

\com{There is a paucity of work on the analysis of extremes with a focus on handling missing data. Even in a general statistical analysis with missing data, it is important to investigate the missingness mechanism as different mechanisms may have different impacts on inference. In general, data with MCAR and MAR missingness mechanisms lead to less precision, but not necessarily bias, whereas MNAR is known to additionally lead to bias in estimation. In the case of an extreme value analysis with the block maxima approach, all three missingness mechanisms can lead to bias as observed block maxima are being used to fit a model when they may not be the true block maxima, hence conflicting with an assumption of the block maxima being identically distributed.}
\

\com{To attempt to avoid such bias, one could drop block maxima from the analysis which were calculated from blocks with sufficiently concerning levels of missing data, however it is generally viewed to be best practice to include all informative data.} To address \com{these issues}, we proposed a variety of different likelihood-based approaches \com{inspired by a right-censoring perspective} in a block maxima framework. \com{The relative preference of the proposed method is shown to depend on the missing data mechanism via an extensive simulation study. Specifically, it was shown that when data are MCAR or MAR, the conditional soft-censoring estimator tends to yield the best results whereas when the larger order statistics were missing from within blocks in the MNAR scenario, the EM method was preferred. The hard censoring approach was shown to be too strict in the simulations and data analysis and can lead to imprecise and grossly inflated return level estimates, particularly in cases with small numbers of blocks. It is also important to note that the simulation scenarios considered are not the only possible ways in which the three missingness mechanisms can appear in real data and so relative performance could differ depending on the exact nature of the missingness mechanism.}
\

Although the proposed estimation procedures are easily implementable in their current state (see supplementary materials) and yield enhanced estimation compared to a standard observed likelihood approach, there remain possible avenues for \com{improving the handling of missing observations within an extreme value analysis. For example, with knowledge of the missingness mechanism, the choice of censoring approach appears more straightforward, however, understanding the nature of the missingness mechanism is not necessarily a straightforward problem. Little's test has been widely adopted to test whether the mechanism is MCAR, but no widely-adopted test exists for distinguishing between MAR and MNAR. For this reason, extending the proposed right-censoring approaches to incorporate additional data sources or variables that help explain the missing data mechanism within a censored likelihood procedure would be of great value.} \comR{It remains an area of future research to develop an improved method of calculating uncertainty and confidence intervals for return levels. With missing data and increased uncertainty, the estimation of high quantiles is a challenging problem. In this work, the empirical coverage of the bootstrap confidence intervals for the proposed approaches depends on the missingness mechanism and percentage of missingness. That is, the coverages are tied to the changes in performance of return level estimation (bias and variance) across varying scenarios. It is unclear how a bootstrapping procedure can be used to incorporate the missingness mechanisms at the series level while also capturing the variability of the block maxima estimators.}
\com{In addition, the peaks over threshold (POT) approach is an alternative method of characterizing extremes to the block maxima approach which warrants an in-depth investigation for how estimation should be modified in the presence of missing data. Unfortunately, there is
no direct analog for the right-censoring approach of this work to be applied to POT as there is no clear case of right censoring as an unobserved threshold exceedance corresponds to an empty record. Finally, it is often of interest to model extremes at multiple locations simultaneously while accounting for missing data. In environmental applications, there are usually spatial dependence patterns that can be exploited through careful modelling to borrow information accross space, e.g., locations that are closer geographically are more likely to report similar extreme surges.} While approaches have been examined in the literature for modelling spatial extremes, an extension which incorporates missing data requires further investigation. 
\\ \vspace{0.5cm}

\noindent{\textbf{Data availability statement}}\\ 
\com{Supplementary materials are available upon request.}


\noindent{\textbf{Data availability statement}}\\
The data analyzed in this paper is publicly available at \url{https://www.tides.gc.ca/en/stations/}.  

\noindent{\textbf{Funding Statement}}\\
This work was supported by two Discovery Grants from the Natural Sciences and Engineering Research Council of Canada.

\noindent{\textbf{Conflict of interest disclosure}}\\
Both authors have no conflicts of interest to declare.

\noindent{\textbf{Ethical approval statement}}\\
Not applicable.

\begin{appendix}

\section{Probabilistic comparisons between observed/unobserved maxima}

%

\begin{theorem}
Let $M_{n_j}$, $M_{n_j'}$ and $M_{n_j + n_j'}$ denote the observed, unobserved and complete data maxima calculated from a random sample of non-negative random variables with cdf $F(\cdot)$ and density function $f(\cdot)$. Then, 
\begin{enumerate}
\item[(i)] $\mathbb{E}(M_{n_j}) \leq \mathbb{E}(M_{n_j + n_j'}) \leq \mathbb{E}(M_{n_j}) + \mathbb{E}(M_{n_j'})$
\item[(ii)] $\mathbb{V}(M_{n_j}) - \mathbb{E}(M_{n_j'}) \left[2 \mathbb{E}(M_{n_j}) + \mathbb{E}(M_{n_j'}) \right] \leq \mathbb{V}(M_{n_j + n_j'}) \leq \mathbb{V}(M_{n_j}) + \mathbb{E}(M^{2}_{n_j'})$
\end{enumerate}
\label{thm1}
\end{theorem}

\begin{proof}
\begin{enumerate}
\item[(i)] The first inequality is established directly:  
$$\mathbb{E}(M_{n_j}) = \int_{0}^{\infty} (1 - F(x))^{n_j} dx \leq \int_{0}^{\infty} (1 - F(x))^{n_j+n_j'} dx = \mathbb{E}(M_{n_j+n_j'})$$
Similarly, the expectation of $M_{n_j+n_j'}$ can be expressed as:  
$$\mathbb{E}(M_{n_j+n_j'}) = \int_{0}^{\infty} n_jx(1 - F(x))^{n_j+n_j'-1} f(x) dx + \int_{0}^{\infty} n_j' x (1 - F(x))^{n_j+n_j'-1}f(x) dx$$
$$ \leq \int_{0}^{\infty} n_jx (1 - F(x))^{n_j-1} f(x) dx + \int_{0}^{\infty} n_j'x (1 - F(x))^{n_j'-1} f(x) dx = \mathbb{E}(M_{n_j}) + \mathbb{E}(M_{n_j'})$$
\item [(ii)] To obtain a lower bound for the variance of $M_{n_j+n_j'}$, an identical argument is used as in part (i) by bounding the variance from below by the second moment of $M_{n_j}$ and the upper bound on $\mathbb{E}(M_{n_j+n_j'})$: 
$$\mathbb{V}(M_{n_j+n_j'}) = \mathbb{E}(M_{n_j+n_j'}^2) - \left(\mathbb{E}(M_{n_j+n_j'})\right)^2 \geq \mathbb{E}(M_{n_j}^2) - \left( \mathbb{E}(M_{n_j}) + \mathbb{E}(M_{n_j'}) \right)^2$$
$$ = \mathbb{V}(M_{n_j}) - \mathbb{E}(M_{n_j'}) \left(2\mathbb{E}(M_{n_j}) + \mathbb{E}(M_{n_j'})\right)$$
To find the upper bound, we use the fact that the second moment of the maximum of the complete sample is bounded from above by the sum of the second moments of the observed and unobserved samples. Since the random variables are non-negative, the lower bound on $\mathbb{E}(M_{n_j+n_j'})$ provides an upper bound for $-(\mathbb{E}(M_{n_j+n_j'}))^2$ yielding:
$$\mathbb{V}(M_{n_j+n_j'}) = \mathbb{E}(M^2_{n_j+n_j'}) - \left( \mathbb{E}(M_{n_j+n_j'}) \right)^2 \leq \mathbb{E}(M_{n_j}^2) + \mathbb{E}(M_{n_j'}^{2}) - \left( \mathbb{E}(M_{n_j}) \right)^2$$
$$ = \mathbb{V}(M_{n_j}) + \mathbb{E}(M_{n_j'}^{2})$$
\end{enumerate}
\end{proof}

\begin{theorem}
Suppose $X_1, ..., X_{n_j}, X'_{1}, ..., X'_{n_j'}$ are a random sample of non-negative random variables with cdf $F(\cdot)$ and pdf $f(\cdot)$. Let $M_{n_j} = \max(X_1, ..., X_{n_j})$ and let $m_{n_j}$ denote its observed value. Then, 
\begin{enumerate} 
\item[(i)] $\mathbb{P}(\exists l \in \{1, ..., n_j'\}: X_l' > m_{n_j}) = 1 - (F(m_{n_j}))^{n_j'}$
\item[(ii)] $\mathbb{P}(\exists l \in \{1, ..., n_j'\}: X_l' > M_{n_j}) = \frac{n_j'}{n_j + n_j'}$
\end{enumerate}
\label{thm2}
\end{theorem}

\begin{proof}
\begin{enumerate}
\item[(i)] The probability that at least one $X_j' > m_{n_j}$ is given by:
$$1 - \mathbb{P}(X_1' \leq m_{n_j}, ..., X_{m_j}' \leq m_{n_j}) = 1 - (F(m_{n_j}))^{n_j'}$$
establishing (i). 
\item[(ii)] The probability statement can be re-expressed as:
$$1 - \mathbb{P}(X_1' \leq M_{n_j}, ..., X_{n_j'}' \leq M_{n_j})$$
$$= 1 - \int_{0}^{\infty} \mathbb{P}(X_1' \leq M_{n_j}, ..., X_{n_j'}' \leq M_{n_j} \mid M_{n_j} = x) n_j f(x) (F(x))^{n_j-1} dx$$
As the variables $X_1', ..., X_{n_j'}'$ are independent of $M_{n_j}$, the above simplifies to:
$$ = 1 - \int_{0}^{\infty} \mathbb{P}(X_1' \leq x, ..., X_{n_j}' \leq x) n_j f(x) (F(x))^{n_j-1} dx$$
$$ = 1 - \int_{0}^{\infty} (F(x))^{n_j'} n_j f(x) (F(x))^{n_j-1} dx = 1 - \int_{0}^{\infty} n_j f(x) (F(x))^{n_j + n_j' - 1} dx$$
$$ = 1 - \frac{n_j}{n_j+n_j'} = \frac{n_j'}{n_j+n_j'}$$
\end{enumerate}
\end{proof}

\end{appendix}
\bibliographystyle{chicago}
\bibliography{sn-bibliography}

\begin{table}[!htbp] \centering \scriptsize
  \caption{\com{Average 50-year return levels with standard errors in parentheses for the proposed modelling approaches computed over 1000 replications (“obs” - observed likelihood, “hard” - right-censored likelihood, “softUC” - unconditional soft-censoring likelihood, “softC” - conditional soft-censoring likelihood, “EM” - EM algorithm) for data generated from an exponential distribution (rate=1) with a scenario I missingness mechanism with 100 observations within blocks. Simulation parameters: “sims” - number of blocks, “pbm” - proportion of blocks with missing observations, “pm” - expected proportion of missing observations within blocks. The true 50-year return level in this simulation is approximately 8.52.} }
  \label{simtable1} 
\begin{tabular}{@{\extracolsep{5pt}} ccc||ccccc} 
\\[-1.8ex]\hline 
sims & pbm & pm & obs & hard & softUC & softC & EM \\ 
\hline \\[-1.8ex]\hline 
 25 & 0.2 & 0.05 & 8.427 (1.359) & 10.309 (4.875) & 8.471 (1.386) & 8.431 (1.356) & 8.64 (1.596) \\ 
 50 & 0.2 & 0.05 & 8.441 (0.855) & 9.743 (1.888) & 8.482 (0.873) & 8.452 (0.86) & 8.655 (0.991) \\ 
100 & 0.2 & 0.05 & 8.476 (0.566) & 9.586 (1.09) & 8.516 (0.577) & 8.49 (0.569) & 8.693 (0.631) \\ 
 25 & 0.5 & 0.05 & 8.509 (1.424) & 273.375 (4424.917) & 8.633 (1.519) & 8.521 (1.422) & 9.195 (2.204) \\ 
 50 & 0.5 & 0.05 & 8.445 (0.831) & 17.9 (17.335) & 8.553 (0.876) & 8.471 (0.839) & 9.054 (1.196) \\ 
100 & 0.5 & 0.05 & 8.456 (0.57) & 14.337 (5.001) & 8.559 (0.597) & 8.488 (0.576) & 9.085 (0.785) \\ 
 25 & 0.8 & 0.05 & 8.521 (1.282) & 331340.34 (10321708.466) & 8.718 (1.411) & 8.547 (1.29) & 10.192 (3.819) \\ 
 50 & 0.8 & 0.05 & 8.43 (0.818) & 1945.728 (28064.174) & 8.607 (0.889) & 8.472 (0.831) & 9.692 (1.861) \\ 
100 & 0.8 & 0.05 & 8.464 (0.545) & 340.61 (3631.162) & 8.637 (0.59) & 8.518 (0.556) & 9.806 (1.202) \\ 
 25 & 0.2 & 0.20 & 8.457 (1.36) & 10.259 (4.344) & 8.636 (1.494) & 8.48 (1.358) & 8.673 (1.516) \\ 
 50 & 0.2 & 0.20 & 8.438 (0.805) & 9.565 (1.605) & 8.591 (0.864) & 8.486 (0.82) & 8.66 (0.915) \\ 
100 & 0.2 & 0.20 & 8.441 (0.574) & 9.447 (1.002) & 8.594 (0.618) & 8.499 (0.59) & 8.662 (0.647) \\ 
 25 & 0.5 & 0.20 & 8.442 (1.494) & 58.175 (426.32) & 8.946 (1.895) & 8.486 (1.445) & 9.075 (2.185) \\ 
 50 & 0.5 & 0.20 & 8.415 (0.872) & 15.823 (10.695) & 8.864 (1.071) & 8.526 (0.897) & 9.054 (1.214) \\ 
100 & 0.5 & 0.20 & 8.338 (0.554) & 13.053 (3.971) & 8.746 (0.663) & 8.476 (0.59) & 8.976 (0.787) \\ 
 25 & 0.8 & 0.20 & 8.434 (1.453) & 4542.127 (79367.421) & 9.372 (2.253) & 8.558 (1.497) & 10.312 (3.83) \\ 
 50 & 0.8 & 0.20 & 8.351 (0.877) & 626.474 (4842.233) & 9.163 (1.247) & 8.533 (0.933) & 9.67 (1.859) \\ 
100 & 0.8 & 0.20 & 8.28 (0.562) & 129.439 (510.086) & 9.039 (0.785) & 8.503 (0.621) & 9.645 (1.218) \\ 
 25 & 0.2 & 0.35 & 8.418 (1.368) & 9.907 (3.642) & 8.696 (1.581) & 8.479 (1.387) & 8.691 (1.667) \\ 
 50 & 0.2 & 0.35 & 8.404 (0.806) & 9.395 (1.535) & 8.645 (0.909) & 8.508 (0.845) & 8.647 (0.935) \\ 
100 & 0.2 & 0.35 & 8.403 (0.526) & 9.29 (0.883) & 8.644 (0.588) & 8.517 (0.554) & 8.64 (0.596) \\ 
 25 & 0.5 & 0.35 & 8.368 (1.558) & 26.71 (190.717) & 9.188 (2.094) & 8.472 (1.365) & 8.996 (1.965) \\ 
 50 & 0.5 & 0.35 & 8.256 (0.816) & 13.376 (7.008) & 8.965 (1.144) & 8.472 (0.901) & 8.903 (1.219) \\ 
100 & 0.5 & 0.35 & 8.293 (0.569) & 12.407 (3.266) & 9.01 (0.77) & 8.565 (0.64) & 8.977 (0.806) \\ 
 25 & 0.8 & 0.35 & 8.201 (1.331) & 2787.672 (62231.594) & 9.93 (2.73) & 8.447 (1.465) & 9.973 (3.771) \\ 
 50 & 0.8 & 0.35 & 8.181 (0.852) & 274.082 (2345.026) & 9.768 (1.646) & 8.556 (0.997) & 9.686 (2.218) \\ 
100 & 0.8 & 0.35 & 8.153 (0.573) & 66.516 (237.273) & 9.645 (1.059) & 8.58 (0.695) & 9.616 (1.32) \\ 
\hline \\[-1.8ex] 
\end{tabular} 
\end{table} 

\begin{table}[!htbp] \centering \scriptsize
  \caption{\com{Average 50-year return levels with standard errors in parentheses for the proposed modelling approaches computed over 1000 replications (“obs” - observed likelihood, “hard” - right-censored likelihood, “softUC” - unconditional soft-censoring likelihood, “softC” - conditional soft-censoring likelihood, “EM” - EM algorithm) for data generated from an exponential distribution (rate=1) with a scenario II missingness mechanism. Simulation parameters: “sims” - number of blocks, “apm” - average proportion of missingness across the series, “n” - number of observations per block. The true 50-year return level in this simulation is approximately 8.52.} }
  \label{simtable2} 
\begin{tabular}{@{\extracolsep{5pt}} ccc||ccccc} 
\\[-1.8ex]\hline 
sims & apm & n & obs & hard & softUC & softC & EM \\ 
\hline \\[-1.8ex] \hline
 25 & 0.05 &  50 & 7.785 (1.412) & 28.464 (84.45) & 8.014 (1.531) & 7.805 (1.338) & 8.397 (1.989) \\ 
 50 & 0.05 &  50 & 7.77 (0.812) & 15.177 (14.553) & 7.985 (0.899) & 7.834 (0.835) & 8.412 (1.222) \\ 
100 & 0.05 &  50 & 7.743 (0.582) & 12.893 (4.23) & 7.945 (0.641) & 7.816 (0.604) & 8.366 (0.831) \\ 
 25 & 0.15 &  50 & 7.661 (1.334) & 817.843 (9211.733) & 8.311 (1.811) & 7.801 (1.308) & 8.846 (2.498) \\ 
 50 & 0.15 &  50 & 7.546 (0.775) & 61.221 (217.356) & 8.108 (1.006) & 7.763 (0.866) & 8.666 (1.577) \\ 
100 & 0.15 &  50 & 7.597 (0.556) & 30.452 (28.382) & 8.172 (0.715) & 7.848 (0.62) & 8.771 (0.989) \\ 
 25 & 0.25 &  50 & 7.524 (1.205) & 1562.903 (14175.445) & 8.495 (1.826) & 7.883 (1.359) & 9.643 (3.712) \\ 
 50 & 0.25 &  50 & 7.399 (0.782) & 503.09 (6600.996) & 8.296 (1.201) & 7.818 (0.967) & 8.991 (2.043) \\ 
100 & 0.25 &  50 & 7.423 (0.538) & 102.528 (235.232) & 8.299 (0.787) & 7.866 (0.675) & 9.158 (1.482) \\ 
 25 & 0.05 & 100 & 8.48 (1.31) & 97.435 (788.715) & 8.719 (1.491) & 8.502 (1.289) & 9.187 (2.002) \\ 
 50 & 0.05 & 100 & 8.481 (0.872) & 20.315 (21.482) & 8.693 (0.962) & 8.541 (0.891) & 9.204 (1.274) \\ 
100 & 0.05 & 100 & 8.424 (0.57) & 16.715 (7.849) & 8.628 (0.629) & 8.499 (0.59) & 9.188 (0.858) \\ 
 25 & 0.15 & 100 & 8.426 (1.327) & 1375.483 (8911.413) & 9.118 (1.932) & 8.598 (1.387) & 10.045 (3.32) \\ 
 50 & 0.15 & 100 & 8.253 (0.783) & 369.266 (4035.345) & 8.829 (1.022) & 8.472 (0.85) & 9.624 (1.784) \\ 
100 & 0.15 & 100 & 8.277 (0.538) & 59.291 (95.241) & 8.834 (0.692) & 8.522 (0.61) & 9.645 (1.168) \\ 
 25 & 0.25 & 100 & 8.205 (1.17) & 286535.912 (8981546.991) & 9.19 (1.895) & 8.607 (1.387) & 11.231 (4.596) \\ 
 50 & 0.25 & 100 & 8.163 (0.773) & 773.839 (2877.84) & 9.053 (1.144) & 8.598 (0.967) & 10.408 (3.064) \\ 
100 & 0.25 & 100 & 8.081 (0.52) & 526.067 (2094.816) & 8.933 (0.768) & 8.52 (0.653) & 10.1 (1.72) \\ 
\hline \\[-1.8ex] 
\end{tabular} 
\end{table} 

\begin{table}[!htbp] \centering \scriptsize
  \caption{\com{Average 50-year return levels with standard errors in parentheses for the proposed modelling approaches computed over 1000 replications (“obs” - observed likelihood, “hard” - right-censored likelihood, “softUC” - unconditional soft-censoring likelihood, “softC” - conditional soft-censoring likelihood, “EM” - EM algorithm) for data generated from an exponential distribution (rate=1) with a scenario III missingness mechanism with 100 observations within blocks. Simulation parameters: “sims” - number of blocks, “pbm” - proportion of blocks with missing observations, “pm” - expected proportion of missing observations within blocks. The true 50-year return level in this simulation is approximately 8.52.} }
  \label{simtable3} 
\begin{tabular}{@{\extracolsep{5pt}} ccc||ccccc} 
\\[-1.8ex]\hline 
sims & pbm & pm & obs & hard & softUC & softC & EM \\ 
\hline \\[-1.8ex] \hline
 25 & 0.2 & 0.05 & 8.127 (1.166) & 8.565 (1.506) & 8.121 (1.159) & 8.122 (1.148) & 8.556 (1.497) \\ 
 50 & 0.2 & 0.05 & 8.135 (0.806) & 8.494 (0.99) & 8.134 (0.804) & 8.147 (0.802) & 8.482 (0.985) \\ 
100 & 0.2 & 0.05 & 8.229 (0.546) & 8.515 (0.624) & 8.231 (0.545) & 8.248 (0.546) & 8.502 (0.621) \\ 
 25 & 0.5 & 0.05 & 9.094 (3.079) & 8.617 (2.037) & 8.949 (2.689) & 8.559 (2.095) & 8.569 (1.961) \\ 
 50 & 0.5 & 0.05 & 8.615 (1.345) & 8.529 (1.295) & 8.555 (1.302) & 8.387 (1.155) & 8.488 (1.272) \\ 
100 & 0.5 & 0.05 & 8.498 (0.802) & 8.511 (0.812) & 8.454 (0.783) & 8.353 (0.737) & 8.468 (0.795) \\ 
 25 & 0.8 & 0.05 & 7.969 (2.882) & 9.054 (3.415) & 8.077 (2.896) & 7.853 (2.27) & 8.646 (2.65) \\ 
 50 & 0.8 & 0.05 & 7.291 (1.218) & 8.587 (2.241) & 7.418 (1.247) & 7.433 (1.195) & 8.359 (1.923) \\ 
100 & 0.8 & 0.05 & 7.145 (0.76) & 8.613 (1.461) & 7.276 (0.782) & 7.356 (0.789) & 8.455 (1.328) \\ 
 25 & 0.2 & 0.20 & 8.211 (5.239) & 8.319 (1.316) & 7.964 (2.8) & 7.918 (1.041) & 8.319 (1.317) \\ 
 50 & 0.2 & 0.20 & 8.137 (0.785) & 8.517 (0.973) & 8.138 (0.78) & 8.16 (0.775) & 8.517 (0.974) \\ 
100 & 0.2 & 0.20 & 8.302 (0.598) & 8.486 (0.651) & 8.312 (0.597) & 8.237 (0.54) & 8.486 (0.651) \\ 
 25 & 0.5 & 0.20 & 29.918 (38.736) & 8.366 (1.742) & 24.629 (31.631) & 7.943 (1.851) & 8.342 (1.615) \\ 
 50 & 0.5 & 0.20 & 25.475 (15.028) & 8.48 (1.46) & 18.597 (12.822) & 7.989 (0.974) & 8.427 (1.199) \\ 
100 & 0.5 & 0.20 & 23.768 (8.272) & 8.448 (0.822) & 17.138 (8.869) & 8.173 (0.698) & 8.448 (0.823) \\ 
 25 & 0.8 & 0.20 & 13.217 (8.941) & 8.491 (2.716) & 17.196 (11.448) & 15.939 (10.409) & 8.417 (2.449) \\ 
 50 & 0.8 & 0.20 & 9.956 (4.36) & 8.383 (1.843) & 13.113 (5.652) & 16.726 (7.697) & 8.345 (1.791) \\ 
100 & 0.8 & 0.20 & 8.714 (2.247) & 8.448 (2.17) & 11.466 (3.1) & 18.127 (6.04) & 8.372 (1.336) \\ 
 25 & 0.2 & 0.35 & 8.02 (5.055) & 8.351 (1.29) & 7.861 (1.032) & 8.25 (1.214) & 8.351 (1.291) \\ 
 50 & 0.2 & 0.35 & 8.081 (0.779) & 8.396 (0.909) & 8.09 (0.773) & 8.33 (0.852) & 8.395 (0.907) \\ 
100 & 0.2 & 0.35 & 8.359 (0.611) & 8.483 (0.641) & 8.375 (0.609) & 8.442 (0.607) & 8.483 (0.641) \\ 
 25 & 0.5 & 0.35 & 66.435 (81.292) & 8.59 (4.398) & 47.229 (65.24) & 7.911 (1.332) & 8.306 (1.61) \\ 
 50 & 0.5 & 0.35 & 53.133 (37.611) & 8.59 (1.964) & 36.818 (37.511) & 8.097 (0.995) & 8.423 (1.203) \\ 
100 & 0.5 & 0.35 & 50.253 (22.921) & 8.699 (2.421) & 28.996 (26.893) & 8.165 (0.666) & 8.415 (0.81) \\ 
 25 & 0.8 & 0.35 & 20.736 (25.081) & 8.723 (9.929) & 41.008 (35.939) & 10.211 (13.893) & 8.4 (2.294) \\ 
 50 & 0.8 & 0.35 & 11.218 (5.98) & 8.584 (4.819) & 23.485 (12.443) & 9.236 (8.574) & 8.212 (1.624) \\ 
100 & 0.8 & 0.35 & 9.689 (3.78) & 8.668 (3.474) & 20.343 (8.325) & 8.242 (3.612) & 8.341 (1.285) \\ 
\hline \\[-1.8ex] 
\end{tabular} 
\end{table}

\begin{table}[!htbp] \centering \tiny
\caption{Estimated 20-, 50-, and 100-year return levels, associated standard errors \com{and 95\% bootstrap confidence intervals} of surge heights (metres) measured at Saint John, Yarmouth and Port-Aux-Basques locations computed using various procedures (``obs'' - observed likelihood, ``hard'' - right-censored likelihood, ``softUC'' - unconditional soft-right-censoring likelihood, ``softC'' - conditional soft-right-censoring likelihood, ``EM'' - EM algorithm)}  
\begin{tabular}{@{\extracolsep{5pt}} c|ccc|ccc|ccc}
& \multicolumn{3}{c}{Return Levels Estimates} & \multicolumn{3}{c}{Standard Errors} & \multicolumn{3}{c}{\com{95\% Confidence Intervals}} \\ \hline 
Location & \multicolumn{9}{c}{Saint John} \\ \hline
Method & 20 years & 50 years & 100 years & 20 years & 50 years & 100 years & \com{20 years} & \com{50 years} & \com{100 years} \\ \hline
Obs & 1.73 & 2.38 & 3.05 & 0.241 & 0.494 & 0.818 & \com{(1.35, 2.29)} & \com{(1.65, 3.57)} & \com{(1.92, 5.10)} \\
Hard &  6.26 & 11.26 & 17.60 & 137.66 & 1745.41 & 11710.62 & \com{(2.81 25.75)} & \com{(3.95, 72.41)} & \com{(4.97, 159.32)} \\
SoftUC &  1.88 & 2.65 & 3.47 & 0.293 & 0.618 & 1.04 & \com{(1.42, 2.58)} & \com{(1.76, 4.12)} & \com{(2.07, 6.14)} \\
SoftC &  2.95 & 4.56 & 6.38 & 0.752 & 1.75 & 3.22 & \com{(1.90, 4.83)} & \com{(2.46, 9.09)} & \com{(3.06, 14.77)} \\ 
EM &  2.65 & 3.74 & 4.84 & 0.689 & 1.44 & 2.45 & \com{(1.77, 4.37)} & \com{(2.18, 7.61)} & \com{(2.51, 11.86)} \\ \hline \hline

Location & \multicolumn{9}{c}{Yarmouth} \\ \hline
Method & 20 years & 50 years & 100 years & 20 years & 50 years & 100 years \\ \hline
Obs & 1.07 & 1.24 & 1.37 & 0.0929 & 0.158 & 0.231 & \com{(0.90, 1.27)} & \com{(0.96, 1.59)} & \com{(1.00, 1.94)} \\
Hard &  5.83 & 12.09 & 21.18 & 87.55 & 832.63 & 4533.48 & \com{(1.65, 47.94)} & \com{(1.98, 196.73)} & \com{(2.23, 566.54))} \\
SoftUC &  1.12 & 1.32 & 1.49 & 0.128 & 0.231 & 0.352 & \com{(0.91, 1.43)} & \com{(0.97, 1.88)} & \com{(1.02, 2.41))}\\
SoftC &  1.17 & 1.42 & 1.64 & 0.129 & 0.233 & 0.3558 & \com{(0.95, 1.45)} & \com{(1.03, 1.95)} & \com{(1.10, 2.52)} \\ 
EM &  1.42 & 1.75 & 2.05 & 0.262 & 0.501 & 0.803 & \com{(0.99, 2.03)} & \com{(1.03, 3.00)} & \com{(1.08, 4.12)} \\ \hline \hline

Location & \multicolumn{9}{c}{Port-Aux-Basques} \\ \hline
Method & 20 years & 50 years & 100 years & 20 years & 50 years & 100 years \\ \hline
Obs & 0.88 & 0.95 & 1.00 & 0.0289 & 0.0426 & 0.0563 & \com{(0.82, 0.93)} & \com{(0.87, 1.03)} & \com{(0.89, 1.11)} \\
Hard &  1.11 & 1.15 & 1.17 & 4.56 & 15.00 & 37.295 & \com{(0.97, 13.43)} & \com{(0.99, 36.61)} & \com{(1.00, 78.49)} \\
SoftUC &  0.90 & 0.97 & 1.02 & 0.0321 & 0.0488 & 0.0654 & \com{(0.83, 0.95)} & \com{(0.88, 1.06)} & \com{(0.91, 1.16)}\\
SoftC &  0.90 & 0.97 & 1.01 & 0.0311 & 0.0449 & 0.0576 & \com{(0.84, 0.96)} & \com{(0.88, 1.05)} & \com{0.90, 1.13)} \\ 
EM &  0.99 & 1.06 & 1.10 & 0.0510 & 0.0768 & 0.101 & \com{(0.89, 1.07)} & \com{(0.92, 1.20)} & \com{(0.93, 1.29)} \\ \hline \hline

\end{tabular}
\label{locations}
\end{table}

\end{document}